\documentclass[reprint,superscriptaddress,amsmath,amssymb,aps,prb]{revtex4-2}
\usepackage{xcolor}
\usepackage{graphicx}
\usepackage{dcolumn}
\usepackage{bm}
\usepackage{hyperref}
\usepackage[normalem]{ulem}
\usepackage[mathlines]{lineno}

\begin{document}

\title{Magnetic anisotropy effect on stabilizing magnetization plateaus of kagome strip chain Heisenberg antiferromagnets}

\author{Chiara Bruzzi}
\affiliation{Department of Material Science and Engineering, University of Pennsylvania, Philadelphia, Pennsylvania 19104, USA}
 
\author{Jian-Xin Zhu}
\affiliation{Theoretical Division, Los Alamos National Laboratory, Los Alamos, New Mexico 87545, USA}
\affiliation{Center for Integrated Nanotechnologies, Los Alamos National Laboratory, Los Alamos, New Mexico 87545, USA}

\author{Yixuan Huang}
\affiliation{Theoretical Division, Los Alamos National Laboratory, Los Alamos, New Mexico 87545, USA}
\affiliation{Center for Integrated Nanotechnologies, Los Alamos National Laboratory, Los Alamos, New Mexico 87545, USA}
\affiliation{Computational Quantum Matter Research Team, RIKEN Center for Emergent Matter Science (CEMS), Wako, Saitama 351-0198, Japan}

\date{\today}

\begin{abstract}
We investigate the anisotropic effect of magnetization plateaus in the antiferromagnetic Heisenberg model on a kagome strip chain. The kagome strip chain Heisenberg model, composed of a hexagonal net of triangles forming five-site unit cells, exhibits four magnetization plateaus in the presence of an applied magnetic field. Using numerical density matrix renormalization group method, we find that the magnetization plateaus are stable against anisotropic interactions in the same direction of the applied magnetic field, but the plateaus become much smaller with anisotropic interactions in other directions. We further show the anisotropic effect of the magnon excitations of the 0.6 plateau state using linear spin wave theory. The magnon bandwidth remains small when tuning the anisotropic interactions along the field where the magnetization plateau is stable, while the band becomes more dispersive with anisotropic interactions perpendicular to the field. In addition, upon tuning down the interaction strength for the two lower legs below a critical value, the Hamiltonian of the kagome strip chain is dominated by two separate spin chains. This can be used to determine the effective lattice structure in materials with strong distortions. Our results enhance the theoretical understanding of the anisotropic effect and the nature of magnetization plateaus in kagome strip chain materials, which can contribute to the design and manipulation of kagome materials with tailored properties.
\end{abstract}


\maketitle

\section{Introduction}
Magnetization plateaus are some of the most striking manifestations of quantum magnetism in low-dimensional spin systems~\cite{sachdev2008quantum, vasiliev2018milestones}. The interplay between strong correlations and frustrated interactions in these magnetic systems can give rise to various exotic many-body states at low temperatures, including magnetization plateaus and even quantum spin liquids~\cite{balents2010spin}. 

As one of the typical frustrated two-dimensional (2D) lattices, the kagome lattice has been of recent interest because of the flat-band-induced strongly-correlated phases as the kagome geometry naturally favors a flat band~\cite{lin2018flatbands, yin2022topological}. Upon a high magnetic field, experiments have found plateaus in the magnetization of the 
spin-$\frac{1}{2}$ kagome  antifferomagnets~\cite{okamoto2011magnetization,goto2016various,okuma2020magnetization,yoshida2022frustrated,yadav2025magnetism,jeon2024one} while numerical calculations on the spin-$\frac{1}{2}$ kagome Heisenberg model have shown magnetization plateaus at $\frac{1}{9},\; \frac{1}{3},\; \frac{5}{9}$ and $\frac{7}{9}$ magnetizations~\cite{nishimoto2013controlling, picot2016spin}, where the nature of the ground state remains hotly debated~\cite{hida2001magnetization,honecker2004magnetization,sakai2011critical, capponi2013numerical, nishimoto2013controlling,capponi2013numerical, picot2016spin,fang2023nature,he2024variational}. Furthermore, different magnetic behaviors under a field perpendicular or parallel to the kagome plane indicate anisotropic spin exchanges~\cite{goto2016various,ihara2020anisotropic}, which usually originate from the strong spin-orbit couplings of the rare-earth ions in kagome materials and may also contribute to the stability of the plateaus~\cite{goto2017ising}. However, theoretical studies of the effect of the magnetic anisotropy in the kagome Heisenberg model are mainly focused on the zero field~\cite{essafi2017generic,seshadri2018topological,benton2021ordered,gomez2024chiral}. Thus, whether the field-induced magnetization plateaus can be stabilized with a strong anisotropic interaction remains an open question.

While the anisotropic effect and the nature of the ground state of magnetization plateaus in a 2D kagome lattice model are far from clear, different types of translational symmetry breaking valence bond crystals~\cite{syromyatnikov2002hidden, singh2007ground, hwang2011spin} have been found in the quasi-1D kagome strip chain at high magnetic fields~\cite{morita2018magnetization, morita2021magnetic}. It would therefore be desirable to start from a kagome strip chain that contains a basic kagome structure and provide a theoretical understanding of the anisotropic effect of the field-induced magnetization plateaus.

The ground state of the isotropic Heisenberg model on a spin-$\frac{1}{2}$ kagome strip chain has been studied to realize a strongly localized Majorana fermion at zero field~\cite{azaria1998kagome,pati1999gapless,ghosh2025frustrated}, and several magnetization plateaus at high fields~\cite{morita2018magnetization}. The magnetization plateaus emerge along with different types of valence bond crystals. More plateaus with larger unit cells~\cite{morita2018magnetization,acevedo2019magnon,morita2021magnetic} and even spin liquids~\cite{morita2021resonating} can be stabilized by tuning the ratio between the coupling strengths in different bonds. However, the anisotropic behavior of the magnetization plateaus has not been fully explored. On the experimental side, $\text{A}_{2}\text{Cu}_{5}(\text{TeO}_{3})(\text{SO}_{4})_{3}(\text{OH})_{4}$ ($A$ = Na, K)~\cite{tang2016synthesis} has a slightly distorted kagome strip chain lattice structure, where theoretical studies have predicted the existence of magnetization plateaus~\cite{bartolome2022theoretical}. In addition, a capped kagome strip spin-$\frac{1}{2}$ lattice material $(\text{NH}_{4})_{2}\text{Cu}_{9}(\text{TeO}_{3})_{4}(\text{MoO}_{4})_{4}(\text{OH})_{4}$~\cite{zhang2020molybdate} has been synthesized recently, where field-dependent magnetization measurement at low temperatures has revealed a possible 0.6 magnetization plateau. 

Motivated by the experimental realization of these kagome strip chains, we study the ground state of the spin-$\frac{1}{2}$ Heisenberg model on a kagome strip chain in a magnetic field using the numerical density matrix renormalization group (DMRG) methods~\cite{white1992density,white1993density,schollwock2011density}. We focus on the Heisenberg interactions in the anisotropic limit. We identify the suppression of the 0.6 magnetization plateau under anisotropic interactions that have the same direction as the magnetic field, while the magnetization plateaus remain robust when the anisotropy is in the other directions. In addition, we explore the effect of inhomogeneous Heisenberg coupling between different sites, which originates from structural distortion in most kagome materials~\cite{ono2009magnetic,okamoto2012distorted,fujihala2014unconventional,downie2015novel,bodaiji2024six}. Below a critical inhomogeneous interaction strength, the Hamiltonian of the kagome strip chain can be approximated by two separate spin chains, where the magnetization plateau of one sub-chain shows up at a different magnetization level. Furthermore, we explore the anisotropic effect of the magnon dispersions using spin wave theory. The magnon dispersions are obtained based on the 0.6 plateau state, where the bandwidth remains small under anisotropic interactions along the magnetic field, but it becomes much larger for anisotropic interactions that are perpendicular to the field. We also show the anisotropic effect of the magnon dispersions corresponding to the polarized state as a comparison.
Our results provide a more comprehensive understanding of the nature of magnetization plateaus and the behavior of magnetization plateaus with different anisotropic interactions in kagome strip chain systems. The critical interaction strength where the Hamiltonian becomes dominated by two separate spin chains has implications in determining the effective lattice structure in materials with strong distortion.


The remainder of the paper is organized as follows. In Sec.~\ref{model}, we introduce the kagome strip chain model and numerical methods. In Sec.~\ref{DMRG}, we present our numerical findings by focusing on the magnetization plateaus with various anisotropic interactions and inhomogeneous interactions. In Sec.~\ref{Spin_wave}, we explore the anisotropic effect of the magnon dispersions with the spin wave theory and provide theoretical understanding of the magnetization plateaus under anisotropic interactions. Section~\ref{Summary} contains the discussion and summary.

\begin{figure}
\centering
\includegraphics[width=0.7\linewidth]{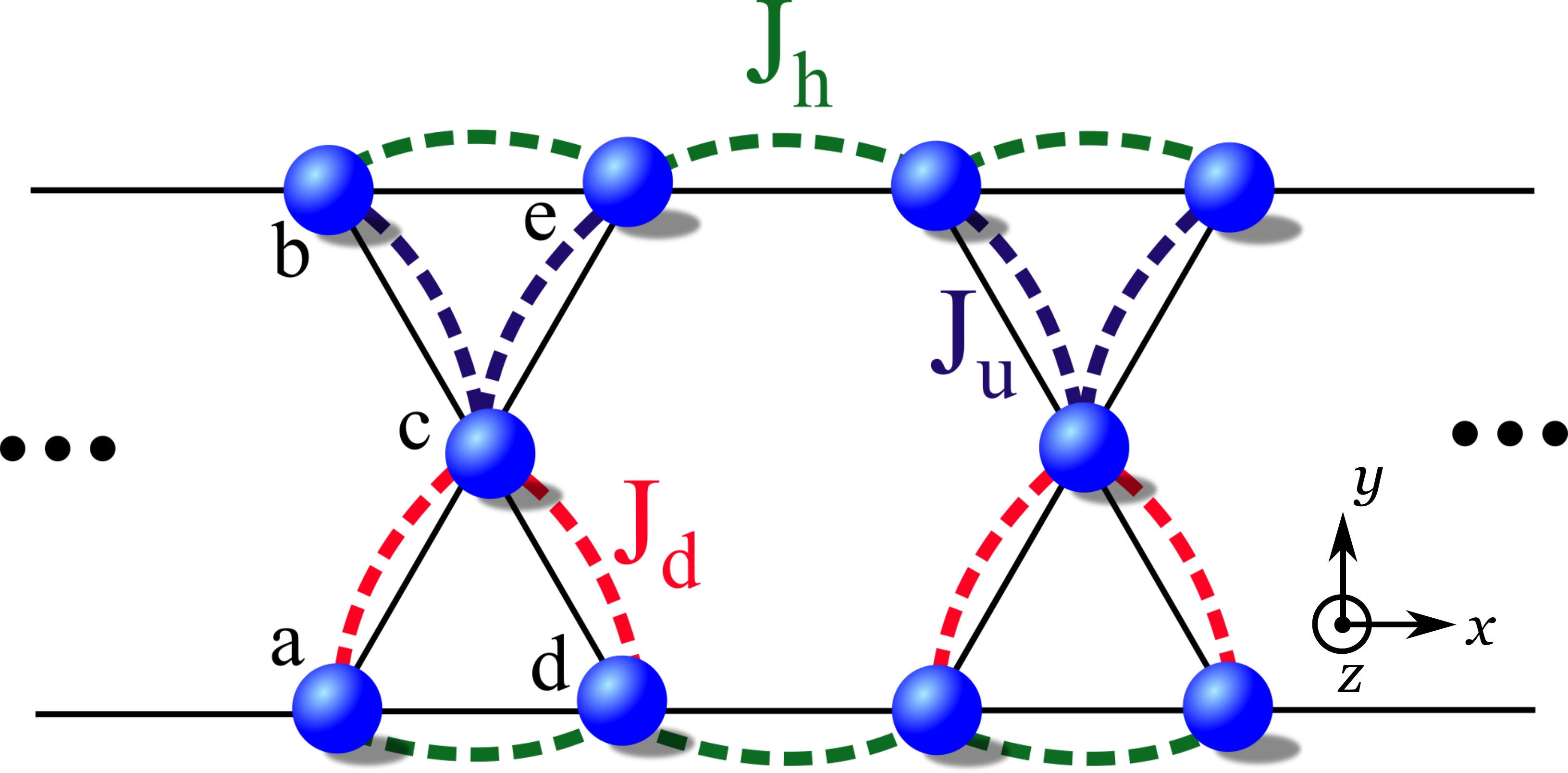}
\caption{Illustration of a kagome strip chain with nearest neighbor Heisenberg interactions. The five sites in the unit cell are labeled as $a$, $b$, $c$, $d$, and $e$.}
\label{Fig1_lattice}
\end{figure}
\begin{figure*}
\centering
\includegraphics[width=0.4\linewidth]{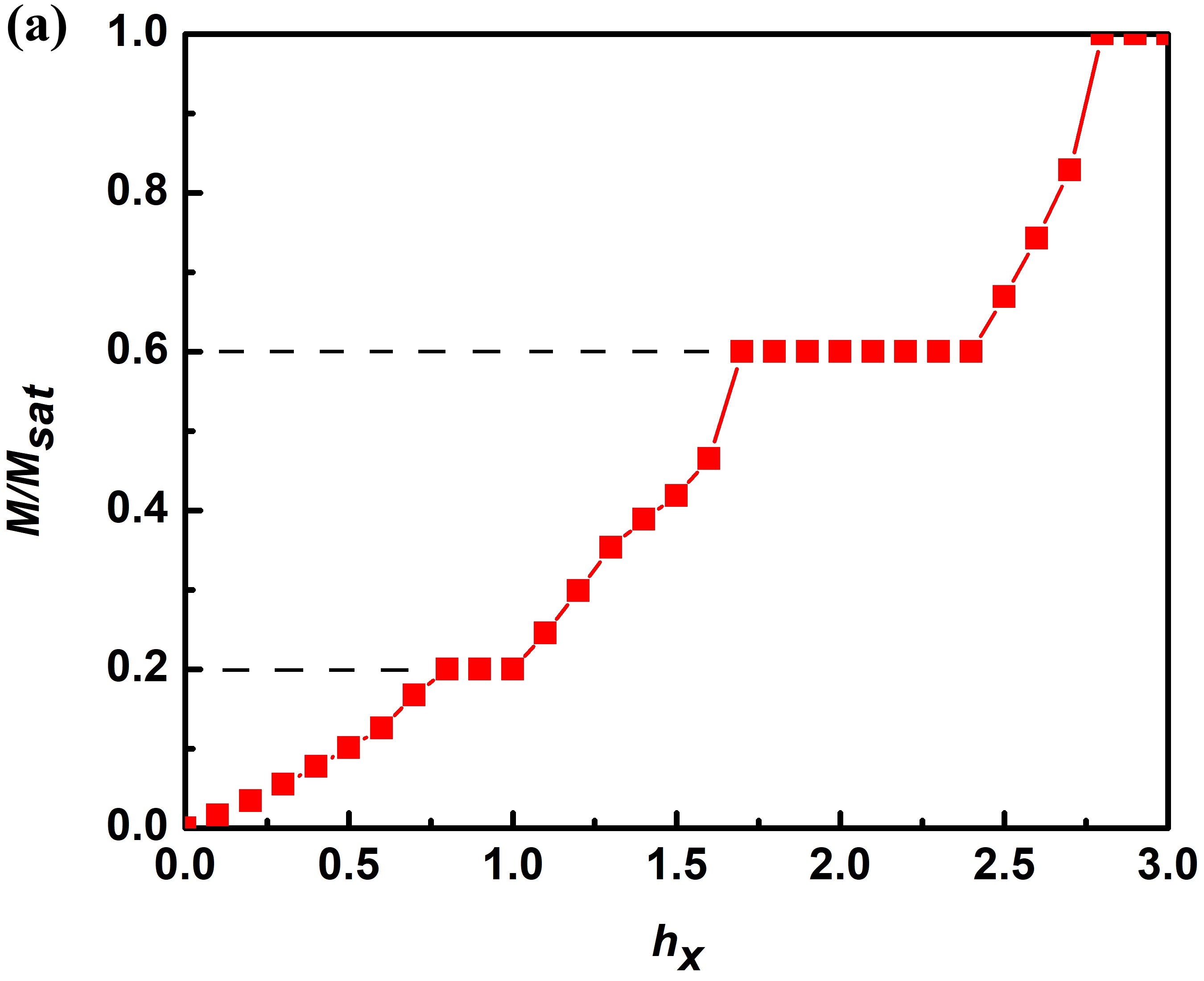}
\includegraphics[width=0.417\linewidth]{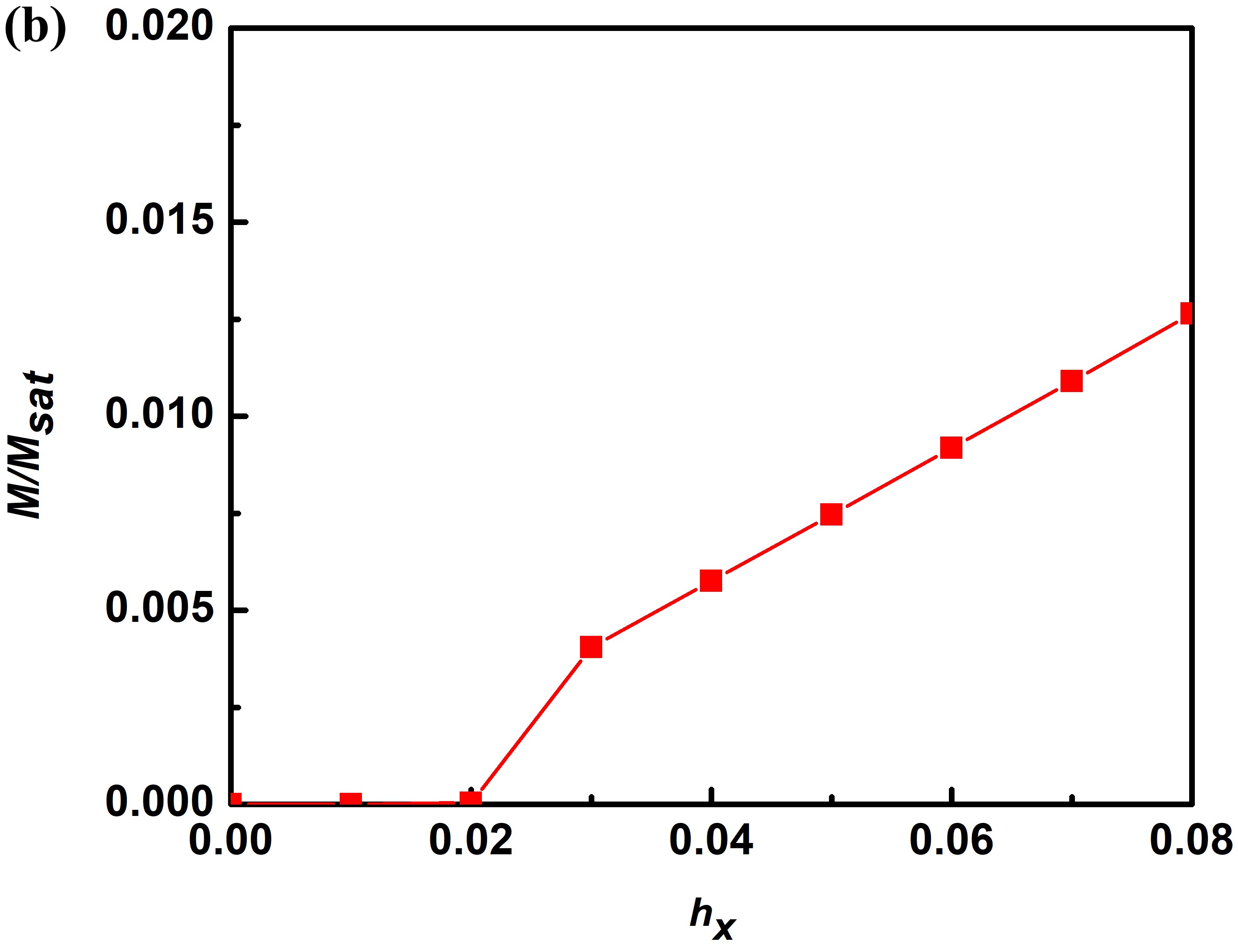}
\includegraphics[width=0.417\linewidth]{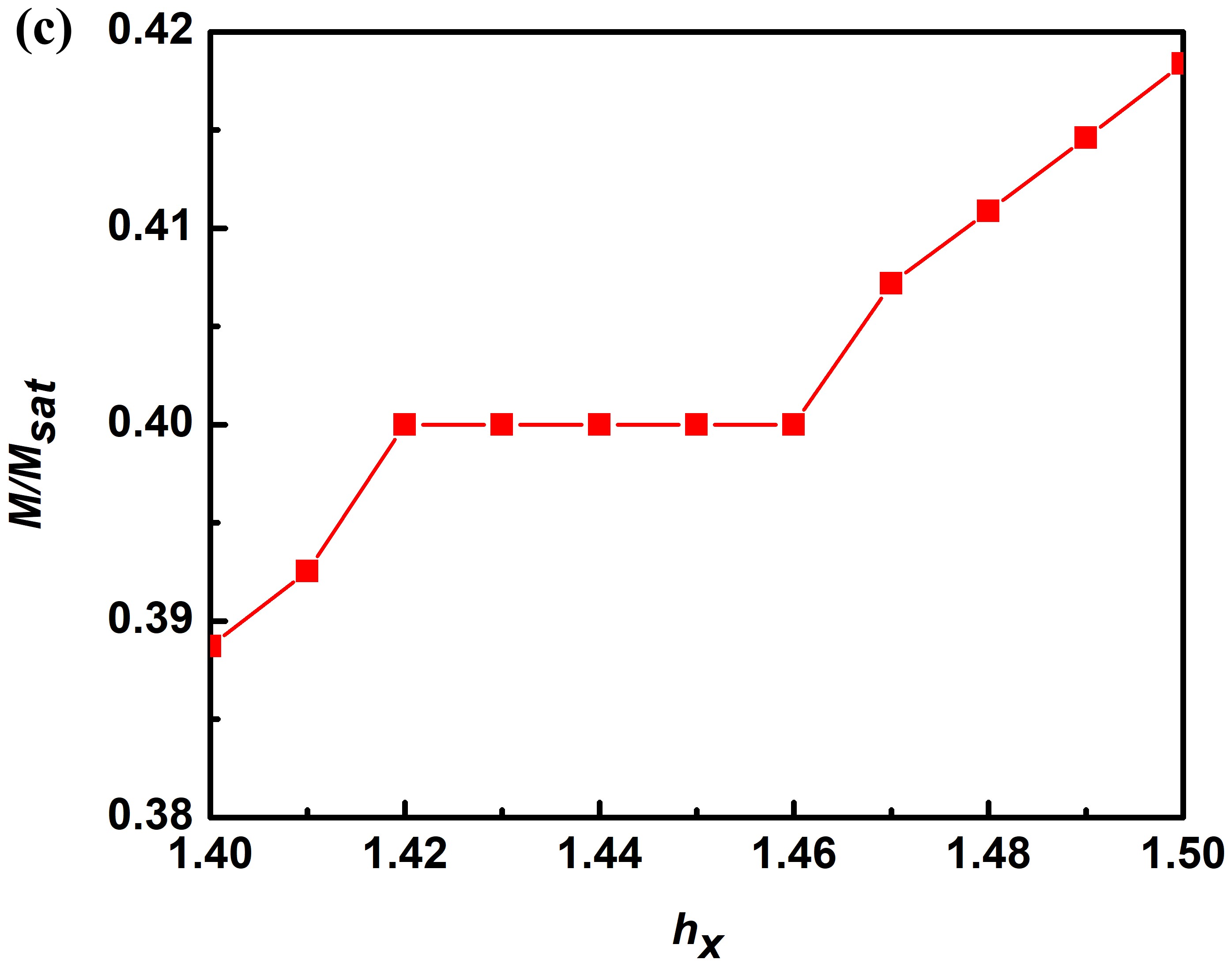}
\includegraphics[width=0.417\linewidth]{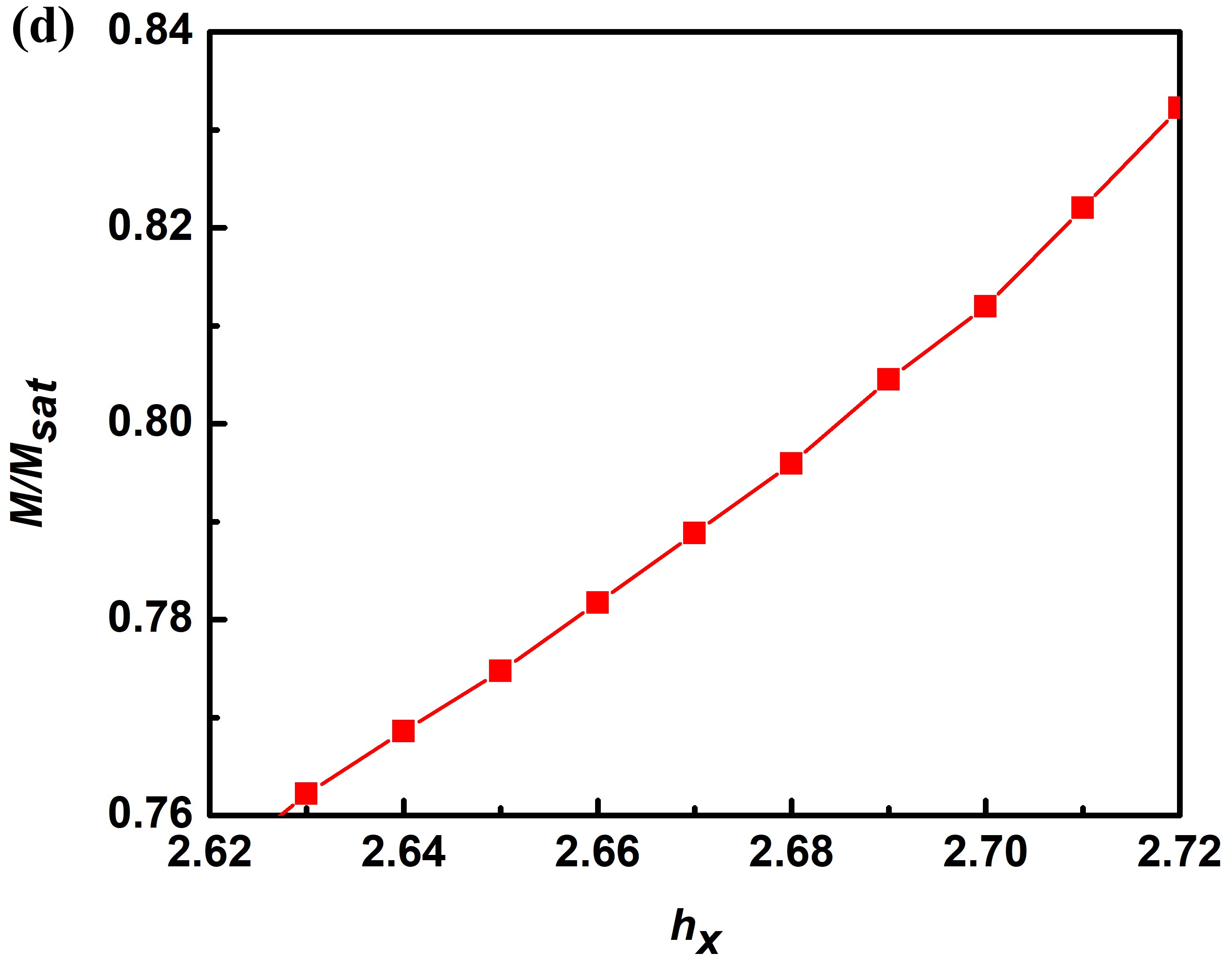}
\caption{The magnetization curve of the isotropic Heisenberg model on a kagome strip chain. Panel (a) shows the curve ranging at $0\leq h_{x} \leq 3$ where the 0.2 and 0.6 magnetization plateaus are identified. Close to saturation, we find that the magnetization increases smoothly over the magnetic field; see more details in the Appendix. Panel (b) shows the curve near the zero magnetization plateau at small $h_{x}$. Panel (c) shows the curve near the 0.4 magnetization plateau. Panel (d) shows the curve near $M/M_{\text{sat}}=0.8$.}
\label{Fig_plateau}
\end{figure*}

\begin{figure}
\centering
\includegraphics[width=0.47\linewidth]{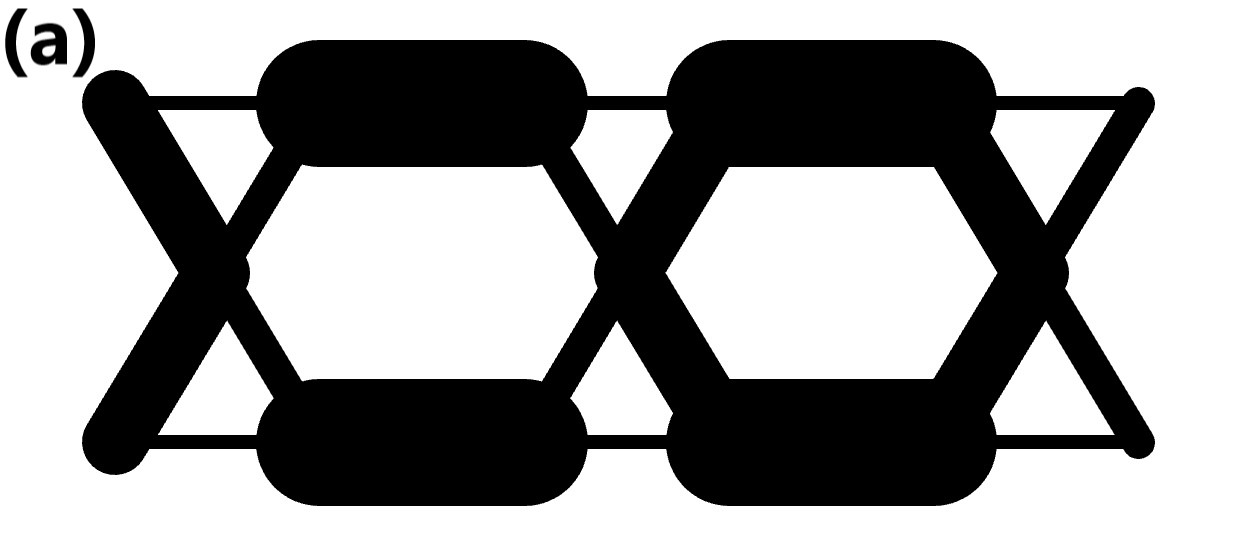}
\includegraphics[width=0.47\linewidth]{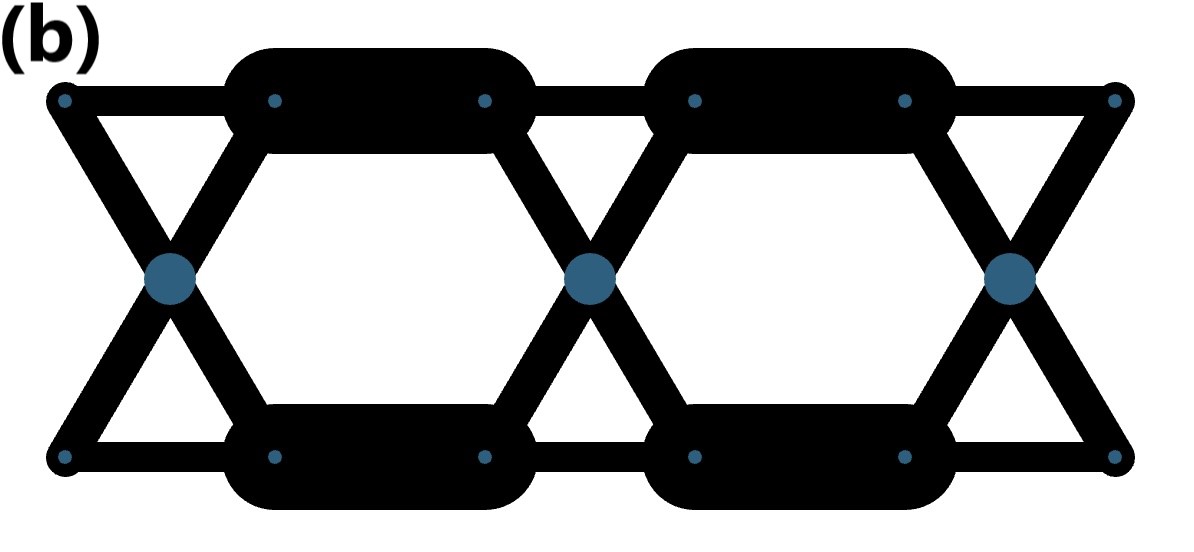}
\includegraphics[width=0.47\linewidth]{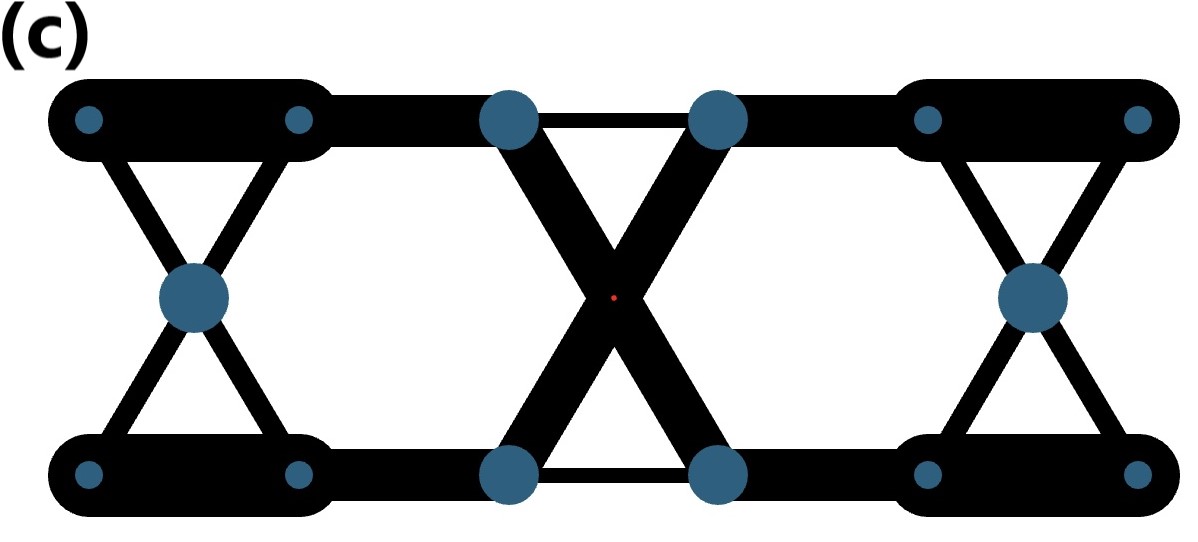}
\includegraphics[width=0.47\linewidth]{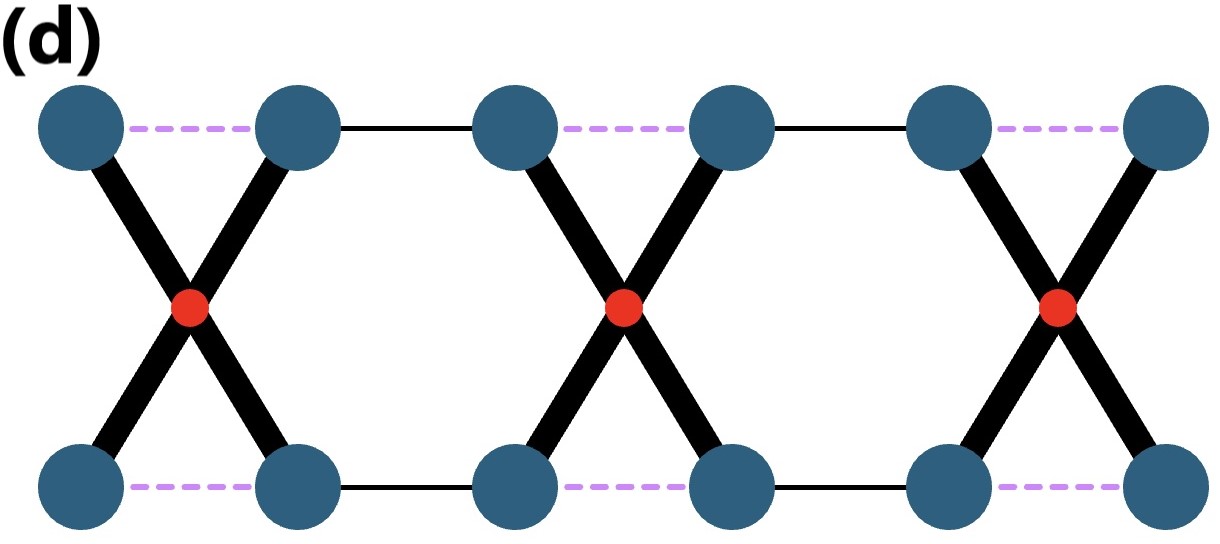}
\caption{The spin correlations and magnetic structures for the four plateaus at (a) $M/M_{\text{sat}}=0$, (b) $M/M_{\text{sat}}=0.2$, (c) $M/M_{\text{sat}}=0.4$, and (d) $M/M_{\text{sat}}=0.6$. Only $5 \times 3$ sites are shown due to the periodic pattern. The purple dashed and black solid lines denote positive and negative values of $\chi(i,j)$ between two neighboring sites, respectively. The line thickness represents the magnitude of $\chi(i,j)$. The dark cyan and red circles respectively denote positive and negative values of onsite spin value $\langle S^{x}_{i} \rangle$ with the diameter representing its magnitude. In panel (d), the $<S_{i}^{x}>$ is -0.18 at the center site $c$, and 0.42 at site $a,b,d,e$. The $\chi(i,j)$ for the bond between site $a$ and site $d$ is 0.06.}
\label{Fig_struc}
\end{figure}

\section{Model and methods}
\label{model}
The kagome strip chain has a five-site unit cell with its geometry similar to the 2D kagome lattice, as shown in Fig.~\ref{Fig1_lattice}. The Hamiltonian of the spin-$\frac{1}{2}$ anisotropic Heisenberg model on the kagome strip chain is defined as

\begin{eqnarray}
\label{eq_H}
H= \sum\limits_{\left\langle ij\right\rangle
}&J_{ij}&(\Delta_{x}S_{i}^{x} S_{j}^{x} + S_{i}^{y} S_{j}^{y} +\Delta_{z} S_{i}^{z} S_{j}^{z}) \nonumber \\
-&h_{x}&\sum\limits_{i}S^{x}_{i},
\end{eqnarray}

where $\langle ij \rangle$ refers to the nearest neighboring sites. To distinguish the coupling strength between different bonds, we label the Heisenberg interactions for the horizontal bonds as $J_{h}$, two upper bonds as $J_{u}$, and two lower bonds as $J_{d}$, as illustrated in Fig.~\ref{Fig1_lattice}. We take $J_{h} = 1$ as the energy unit. $h_{x}$ refers to the magnetic field applied in the $x$ direction. The $\Delta _{x}$ and $\Delta _{z}$ represent the anisotropic Heisenberg interactions (the $x$ direction is taken to be along the chain and $z$ is chosen to point out of the strip chain plane) while $J_{u}$ and $J_{d}$ can be tuned to explore the inhomogeneous effect. 

We apply the infinite DMRG methods to the kagome strip chain consisting of multiples of six unit cells to accommodate the translational symmetry breaking states, unless otherwise specified. Calculations are performed using the TeNPy Library (version 0.10.0)~\cite{tenpy}. We note that the number of total $S^{z}$ sectors are limited by the relatively small unit cell size as compared to the whole lattice size in finite DMRG calculations~\cite{morita2018magnetization}; thus, we specifically choose the magnetic field in the $x$ direction to break the total $S^{z}$ conservation and obtain a smooth magnetization curve. To determine the magnetic plateaus, we directly calculate the magnetization $M=\frac{1}{n} \sum ^{n}_{i=1} \langle S^{x}_{i} \rangle$ in the same direction as the magnetic field. Here $n$ is the total number of sites. We keep up to 800 bond dimensions to reach a numerical truncation error around $1 \times 10^{-7}$. For critical parameter regimes near the phase transitions, we use unit cells up to 12 for better numerical convergence.


\section{Numerical DMRG results}
\label{DMRG}
First, we analyze the isotropic case for the kagome strip chain, where $J_{h}=J_{u}=J_{d}=1$ and $\Delta _{x}=\Delta _{z}=1$. The magnetic field $h_{x}$ is varied by increments of 0.1 for the regime $0\leq h_{x} \leq 3$. The magnetization $\textit{M}$ is evaluated in the same direction as the applied magnetic field and normalized by the saturation magnetization $M_{\text{sat}}=0.5$.

Figure~\ref{Fig_plateau}(a) shows the magnetization curve obtained in the isotropic case. The figure shows two magnetization plateaus at $M/M_{\text{sat}} = 0.2$ and 0.6. The results are consistent with previous studies~\cite{morita2018magnetization}. To identify possibly smaller plateaus that exist in the isotropic limit~\cite{morita2021magnetic}, we adapt a more fine-grained examination of the magnetization curve by varying $h_{x}$ with a smaller step size of 0.01. As shown in Fig.~\ref{Fig_plateau}(b), a zero magnetization plateau appears in the range of $0\leq h_{x} \leq 0.02$ with a very small deviation of $M$ from zero. The deviations are in the order of the truncation error, suggesting that the zero magnetization plateau is within numerical accuracy. Additionally, we identify a plateau in the range of $1.42\leq h_{x} \leq 1.46$ with the corresponding magnetization of $M/M_{\text{sat}} = 0.4$, as shown in Fig.~\ref{Fig_plateau}(c). Comparatively, there is no plateau near $M/M_{\text{sat}} = 0.8$ in the isotropic limit [Fig.~\ref{Fig_plateau}(d)]. These plateaus contribute to magnetic processes of the isotropic kagome strip chain.

\begin{figure}
\centering
\includegraphics[width=0.95\linewidth]{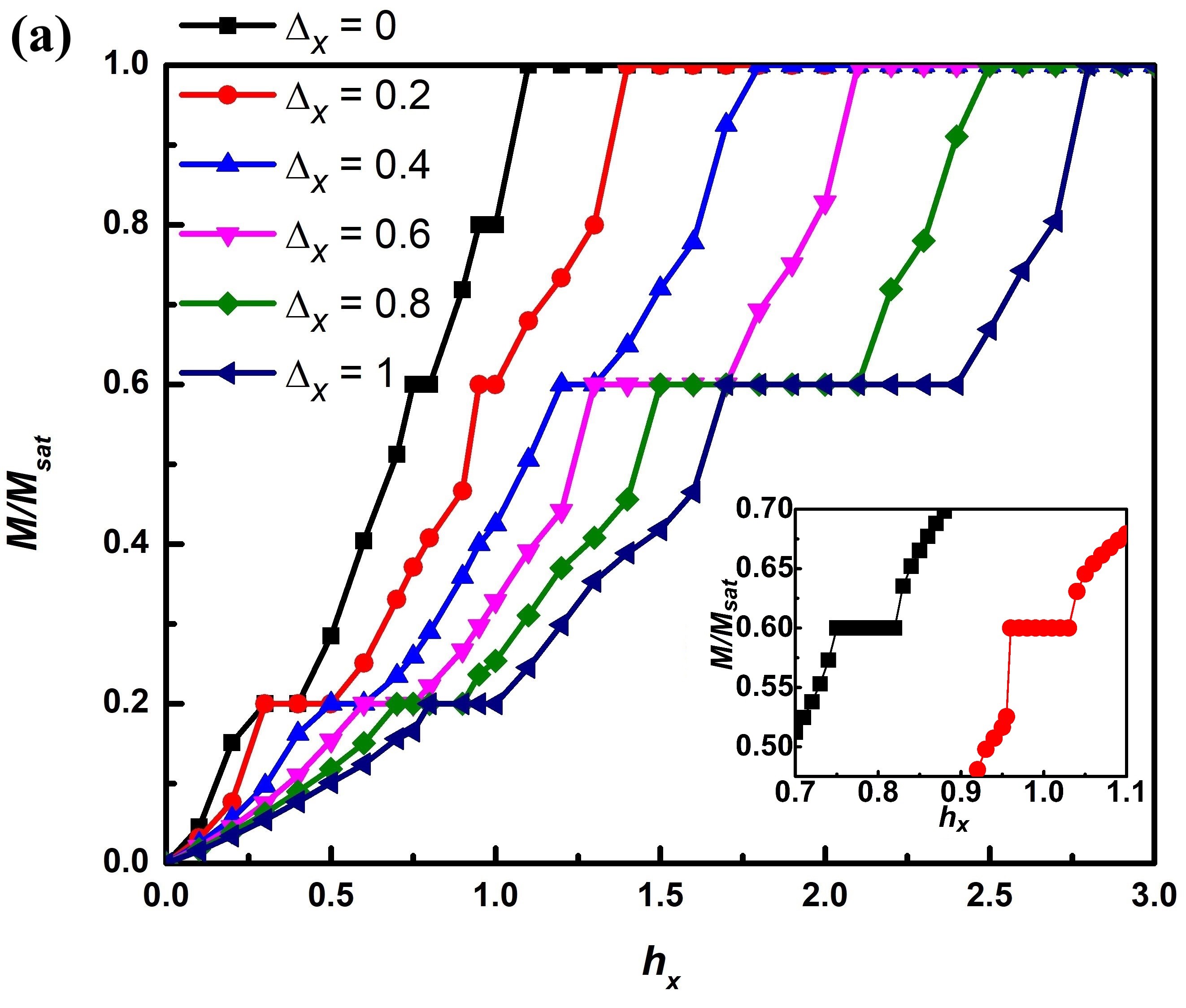}
\includegraphics[width=0.95\linewidth]{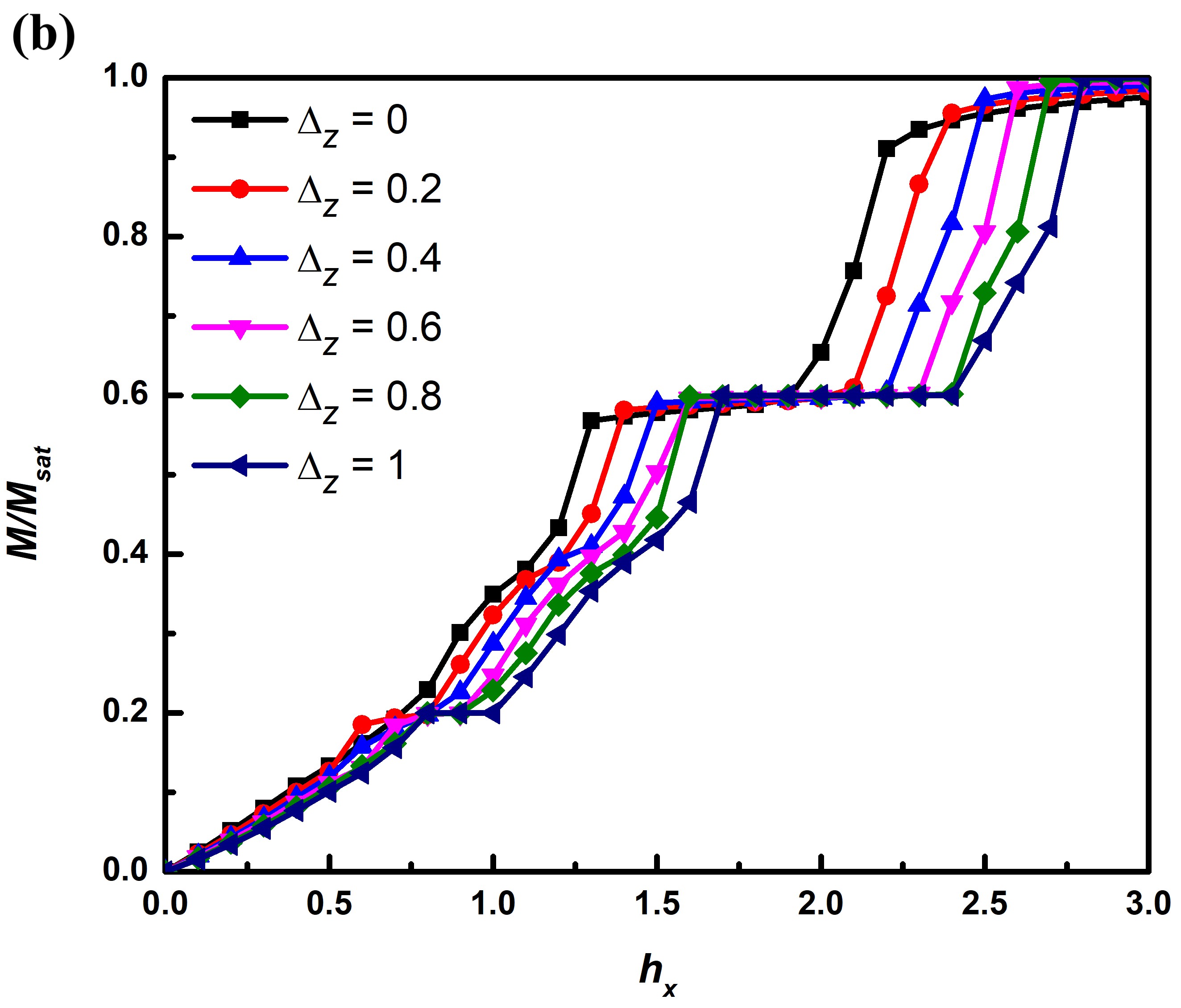}
\caption{The magnetization curve of the anisotropic Heisenberg model with (a) varying $\Delta _{x}$ but fixed $\Delta _{z}=1$ and (b) varying $\Delta _{z}$ but fixed $\Delta _{x}=1$. The inset of panel (a) shows the same 0.6 plateau in a smaller regime. We also identify the emergence of a narrow plateau at $M/M_{\text{sat}} = 7/15$ in the anisotropic limit for $0.2< \Delta _{x}<0.4$; see more details in the Appendix.}
\label{Fig_delta}
\end{figure}

\begin{figure}
\centering
\includegraphics[width=0.95\linewidth]{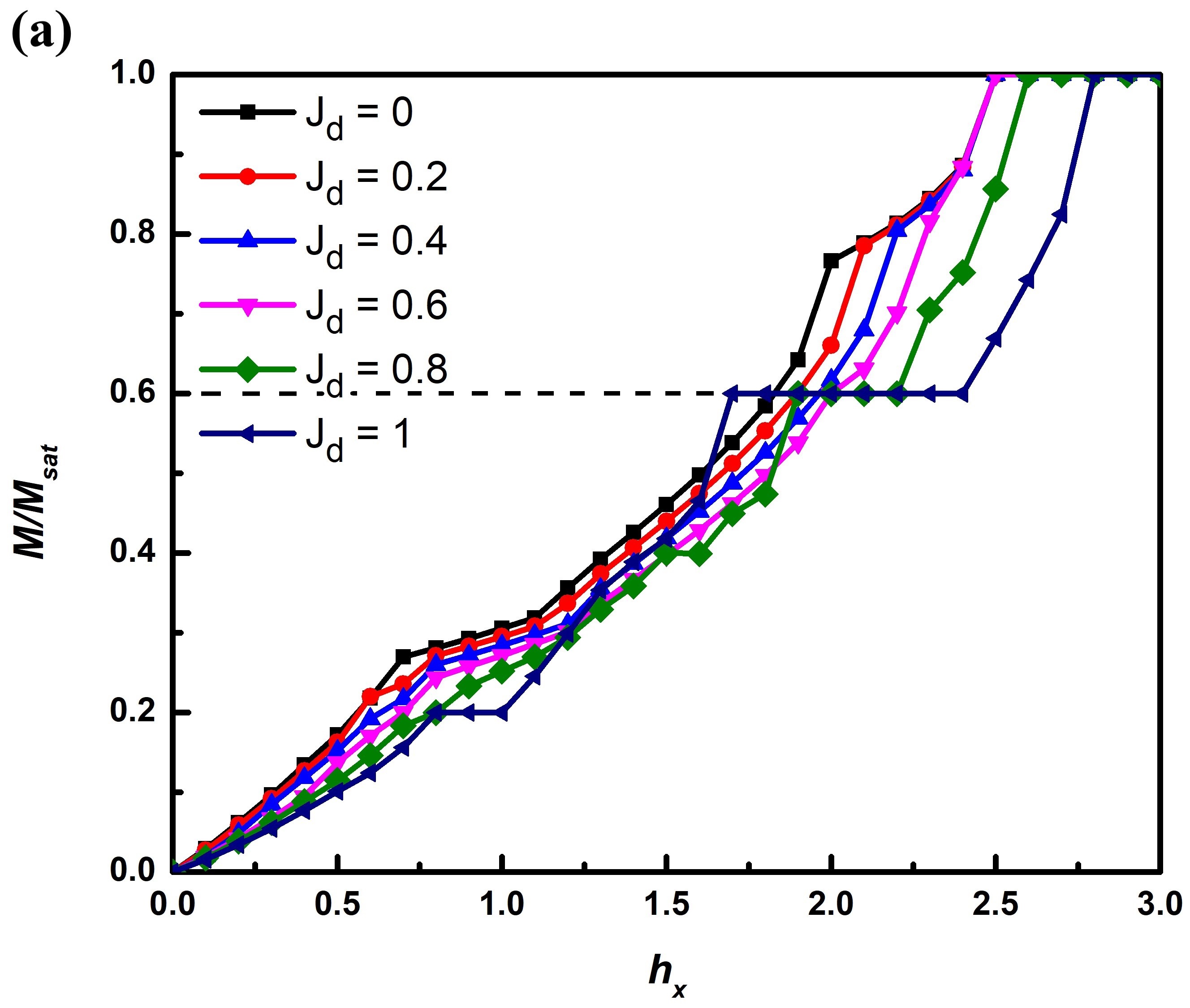}
\includegraphics[width=0.95\linewidth]{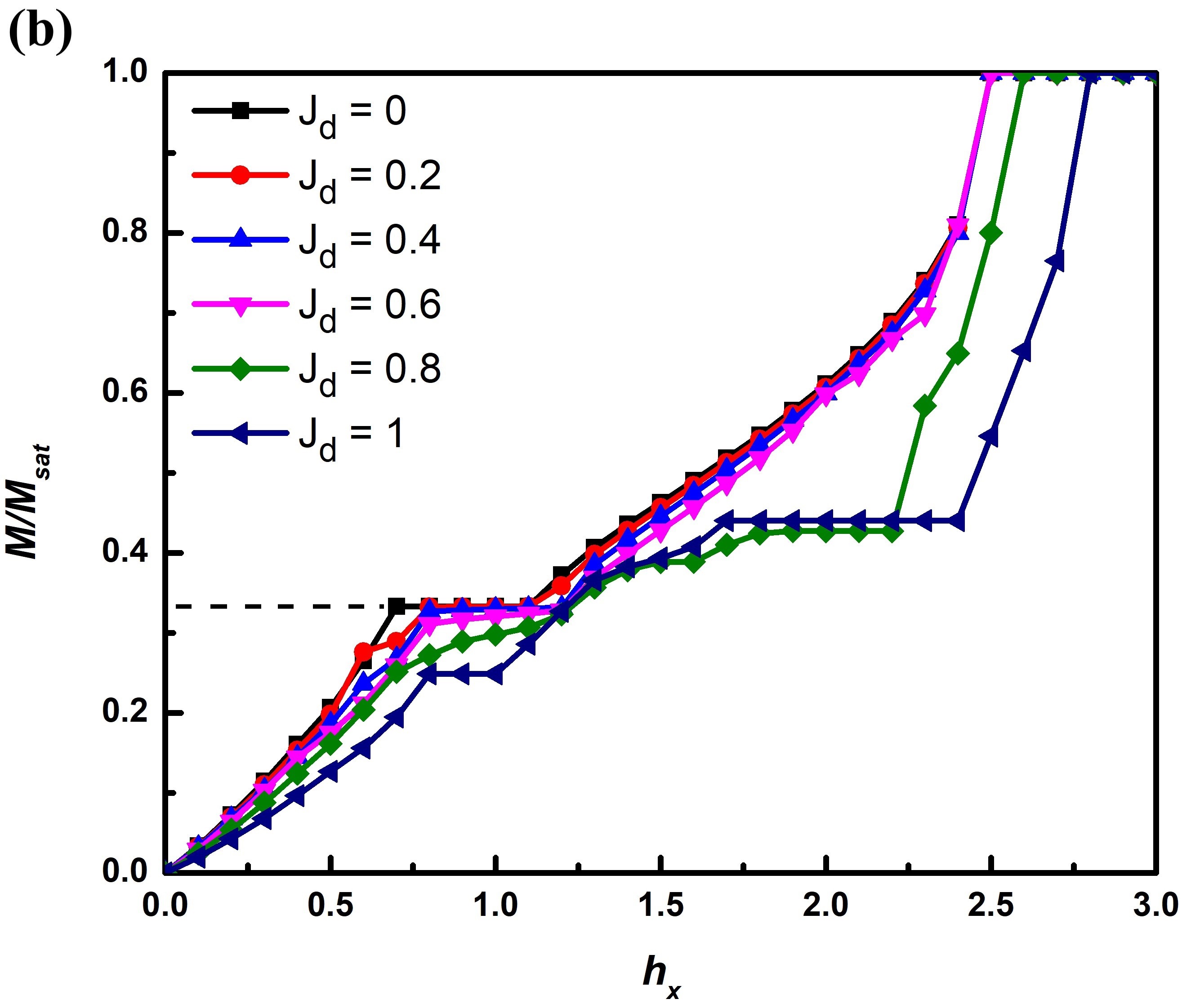}
\caption{The magnetization curve where the magnetization is averaged over (a) all five sites and (b) top three sites with various $J_{d}$ interaction strengths. The plateaus are indicated by the dashed lines.}
\label{Fig_Jd}
\end{figure}

To explore the nature of the plateaus in Fig.~\ref{Fig_plateau}, we analyze the spin correlations between neighboring sites, defined as $\chi(i,j) = \langle \mathbf{S}_i \cdot \mathbf{S}_j \rangle - \langle \mathbf{S}_{i} \rangle \cdot \langle \mathbf{S}_{j} \rangle$. 
Figures~\ref{Fig_struc}(a) and (c) show the spin correlations at different neighboring bonds as well as the onsite magnetization at different sites for $M/M_{\text{sat}}=$ 0 and 0.4 plateaus, respectively. The spin correlation and onsite magnetization pattern breaks translational symmetry with a period of two unit cells. In contrast, the 0.2 and 0.6 magnetization plateaus hold translational symmetry, as shown in Figs.~\ref{Fig_struc}(b) and (d), respectively. The magnetic structures of the ground state at different plateaus indicate valence bond crystal states with a finite spin gap, and the value of the spin gap can be estimated by $\frac{1}{2}(h_{u}-h_{d})$ where $h_{u}$ and $h_{d}$ are the upper and lower critical magnetic fields for each plateau, respectively.

\subsection{Anisotropic Heisenberg interactions}
We then turn to the anisotropic effects of the model on the magnetization plateaus. To see this effect clearly, we focus on the $M/M_{\text{sat}}=0.6$ plateau which is most robust in the isotropic limit. We examine the influence of the anisotropic Heisenberg interactions that are applied along and perpendicular to the chain, denoted by $\Delta_x$ and $\Delta_z$, respectively. The values of $J_{h}$, $J_{u}$, and $J_{d}$ remain equal to 1. First, we keep $\Delta_z = 1$ and vary $\Delta_x$. As shown in Fig.~\ref{Fig_delta}(a), the width of the 0.6 magnetization plateau decreases as $\Delta_x$ decreases from 1 to 0 and the width of the plateau is around 0.07 for $\Delta_x = 0$, as shown in the inset of Fig.~\ref{Fig_delta}(a). Then we vary the $\Delta_z$ parameter while keeping $\Delta_x = 1$. As shown in Fig.~\ref{Fig_delta}(b), the 0.6 magnetization plateau persists for all values in the range $0 \leq  \Delta _{z} \leq 1$ with similar width, suggesting that the plateau is robust against anisotropic interactions in the $z$ direction. The width of the plateau slightly decreases as $\Delta_{z}$ decreases.

Additionally, we observe in Fig.~\ref{Fig_delta}(b) that for a smaller $\Delta_z$, the magnetization of the 0.6 plateau slightly deviates from $M/M_{\text{sat}}=0.6$ over the decrease of the magnetic field. However, for $\Delta_z =1$ the 0.6 plateau is completely flat. The completely flat magnetization plateau is related to the Oshikawa-Yamanaka-Affleck condition~\cite{oshikawa1997magnetization} with spin $U$(1) symmetry, which is $n_{uc}p(M_{sat}-M)=$ integer where $n_{uc}$ is the number of sites in a unit cell and $p$ is the ground state periodicity in terms of the unit cell. All four magnetization plateaus found in Fig.~\ref{Fig_plateau} satisfy the condition. For $\Delta _z < 1$ the spin $U$(1) symmetry is broken in the $x$ direction. Thus, the Oshikawa-Yamanaka-Affleck condition cannot be applied to ensure a completely flat magnetization plateau.

\subsection{Inhomogeneous Heisenberg interactions}
The distorted structures in materials can result in nonequivalent exchange interactions between different sites. If the exchange interactions at certain bonds are weak enough, the Hamiltonian can be approximated by two separate subchains.
To investigate the inhomogeneous effect of the kagome strip chain, we manipulate the $J_d$ parameter while keeping $\Delta_x = \Delta_z =1$ and $J_u = J_h =1$. As shown in Fig.~\ref{Fig_Jd}(a), the 0.6 magnetization plateau remains robust for $J_d>0.8$. Below $J_d=0.6$, the 0.6 magnetization plateau vanishes, indicating a phase transition. The phase transition can be further confirmed with the magnetization curve considering only top three sites in a unit cell [Fig.~\ref{Fig_Jd}(b)], where a new magnetization plateau is identified at $M/M_{\text{sat}} = 1/3$. The two plateaus for $J_{d}=1$ shown in Fig.~\ref{Fig_Jd}(b) correspond to the 0.2 and 0.6 plateaus in Fig.~\ref{Fig_Jd}(a). The 1/3 magnetization plateau is consistent with previous studies on the frustrated antiferromagnetic Heisenberg spin chain~\cite{okunishi2003magnetic,honecker2004magnetization}, suggesting that the Hamiltonian for $J_d<0.6$ is dominated by one chain of frustrated Heisenberg antiferromagnets and one single-leg chain of Heisenberg antiferromagnets with only nearest neighbor interaction.



\section{Linear spin wave theory}
\label{Spin_wave}

To explore the anisotropic effect of the spin excitations of the plateau state, we analyze the dispersion of magnons with a semi-classical magnon model starting from the plateau state~\cite{plat2018kinetic}. First we consider the 0.6 plateau state. The ground state can be approximated by one spin $\downarrow $ at site $c$ in the center of the unit cell, and four spin $\uparrow $ at sites $a$, $b$, $d$, $e$, as illustrated in Fig.~\ref{Fig1_lattice}. As shown in Fig.~\ref{Fig_struc}(d), such spin configuration is consistent with the magnetic structure of the 0.6 plateau state and the correlations between neighboring sites remain weak for $0 \leq \Delta _z \leq 1$. The onsite spin value of the site $c$ becomes closer to -0.5 for decreasing $\Delta _z$, but it increases as $\Delta _x$ decreases. Thus, the spin wave analysis can provide a qualitative understanding of the spin excitations of the 0.6 plateau state for larger $\Delta _x$.

We assign positive spins at sites $a,b,d,e$ as A spins, and negative spin at site $c$ as B spins in the unit cell. The Holstein-Primakoff transformation~\cite{holstein1940field} is used to map the spin operators onto a set of boson creation and annihilation operators as given in Eq.~(\ref{eq_HP}), where $a_{u,v}$ ($b_{u,v^{\prime}}$) is the annihilation operator for A (B) spins. The spin operators $S^{-}_{u,v}$ and boson operators $a_{u,v}$ are labeled by the unit cell index $u$, and the site index $v$ within the unit cell. Here we assume $v\in \{a,b,d,e \}$ and $v^{\prime}\in \{c\}$:


\begin{align}
\label{eq_HP}
&S_{u,v}^{+}=\sqrt{2S-a^{\dagger }_{u,v}a_{u,v}}\;a_{u,v}, \nonumber \\ 
&S_{u,v}^{-}=a_{u,v}^{\dagger }\sqrt{2S-a^{\dagger }_{u,v}a_{u,v}}, \nonumber \\
&S_{u,v}^{z}=S-a^{\dagger }_{u,v}a_{u,v}, \\
&S_{u,v^{\prime}}^{+}=b_{u,v^{\prime}}^{\dagger }\sqrt{2S-b^{\dagger }_{u,v^{\prime}}b_{u,v^{\prime}}},  \nonumber \\
&S_{u,v^{\prime}}^{-}=\sqrt{2S-b^{\dagger }_{u,v^{\prime}}b_{u,v^{\prime}}}\;b_{u,v^{\prime}}, \nonumber \\
&S_{u,v^{\prime}}^{z}=b^{\dagger }_{u,v^{\prime}}b_{u,v^{\prime}}-S. \nonumber
\end{align}

Using the Fourier transformation to the momentum space where the boson operators are defined as $a_{u,v}=\frac{1}{\sqrt{N/5}}\sum_{k}e^{ikx_{u}}a_{k,v}$ and $b_{u,v^{\prime}}=\frac{1}{\sqrt{N/5}}\sum_{k}e^{ikx_{u}}b_{k,v^{\prime}}$, the Hamiltonian can be written as

\begin{align}
\label{eq_vector19}
H=&\sum _{k}\Phi _{k}^{\dagger } [H]_{k} \Phi _{k}, \\
\Phi _{k}^{\dagger }=(a_{k,a}^{\dagger },&a_{k,b}^{\dagger },b_{-k,c},a_{k,d}^{\dagger },a_{k,e}^{\dagger }) \nonumber
\end{align}

where $[H]_{k}$, which corresponds to the 0.6 plateau state, is a $5\times 5$ matrix given by


\begin{widetext}
\begin{align}
\label{eq_Magnon19}
[H]_{k}=\begin{bmatrix}
h_{x} - S\Delta _{x} & 0 & \frac{S}{2}(\Delta _{z} + 1) & \frac{S}{2}(\Delta _{z} + 1)(1+e^{-ik}) & 0 \\
0 & h_{x} - S\Delta _{x} & \frac{S}{2}(\Delta _{z} + 1) & 0 & \frac{S}{2}(\Delta _{z} + 1)(1+e^{-ik}) \\
\frac{S}{2}(\Delta _{z} + 1) & \frac{S}{2}(\Delta _{z} + 1) & -h_{x} + 4S\Delta _{x} & \frac{S}{2}(\Delta _{z} + 1) & \frac{S}{2}(\Delta _{z} + 1) \\
\frac{S}{2}(\Delta _{z} + 1)(1+e^{ik}) & 0 & \frac{S}{2}(\Delta _{z} + 1) & h_{x} - S\Delta _{x} & 0 \\
0 & \frac{S}{2}(\Delta _{z} + 1)(1+e^{ik}) & \frac{S}{2}(\Delta _{z} + 1) & 0 & h_{x} - S\Delta _{x} \\
\end{bmatrix} .
\end{align}

Here we assume all interactions have equal strength $J_{d}=J_{u}=J_{h}=1$ and ignore higher order terms. The lattice constant of the chain is set to 1 for simplicity. The magnon dispersion are obtained through the Bogoliubov transformation where the quasi-particle excitations naturally obey the bosonic commutation relations. For a generic quadratic bosonic Hamiltonian the row vector has ten components containing both annihilation and creation operators

\label{eq_operator}
\begin{align}
\Phi _{k}^{\dagger }&=(a_{k,a}^{\dagger },...b_{k,c}^{\dagger },...a_{k,e}^{\dagger },a_{-k,a},...b_{-k,c},...,a_{-k,e}).
\end{align}

Following Ref.~\cite{smit2020magnon}, the magnon excitations can be obtained by diagonalizing the dynamical matrix which is defined as

\label{eq_dynamical}
\begin{align}
[H]_{k}^{dyn}&= \mathbb{G}[H]_{k}, \\
\mathbb{G}&=\begin{bmatrix}
 \textbf{1} & 0 \\
 0 & -\textbf{1} \\
\end{bmatrix} \nonumber
\end{align}

where $\textbf{1}$ is a five-dimensional identity matrix. We note that the resulting dynamical matrix $[H]_{k}^{dyn}$ is different from $[H]_{k}$, as opposed to the case for Fermions. A general algorithm to obtain the magnon excitations of this type of Hamiltonian is proposed by Colpa~\cite{colpa1978diagonalization}; also see discussions in Ref.~\cite{smit2020magnon} and Refs.~\cite{maldonado1993bogoliubov,serga2012brillouin}.

In our model, the $[H]_{k}$ is reduced to a $5 \times 5$ matrix in Eq.~(\ref{eq_Magnon19}) by getting rid of the zero matrix elements and the corresponding row vector contains only five components, as given in Eq.~(\ref{eq_vector19}). Thus, solving $[H]_{k}^{dyn}$ for the eigenvalues $\varepsilon (k)$ we arrive at five energy bands.

For comparison we turn to the fully polarized state at a sufficiently high magnetic field. The spins at all five sites in the unit cell are assigned as A spins. Similar to Eq.~(\ref{eq_HP}) we use the Holstein-Primakoff transformation followed by the Fourier transformation to obtain the momentum space bosonic Hamiltonian. The resulting Hamiltonian and $[H]_{k}$ is given by Eq.~(\ref{eq_vector3}) and Eq.~(\ref{eq_Magnon3}), respectively,

\begin{align}
\label{eq_vector3}
H=&\sum _{k}\Psi _{k}^{\dagger } [H]_{k} \Psi _{k}, \\
\Psi _{k}^{\dagger }=(a_{k,a}^{\dagger },&a_{k,b}^{\dagger },a_{k,c}^{\dagger },a_{k,d}^{\dagger },a_{k,e}^{\dagger }). \nonumber
\end{align}

\begin{align}
\label{eq_Magnon3}
[H]_{k}=\begin{bmatrix}
h_{x} - 3S\Delta _{x} & 0 & \frac{S}{2}(\Delta _{z} + 1) & \frac{S}{2}(\Delta _{z} + 1)(1+e^{-ik}) & 0 \\
0 & h_{x} - 3S\Delta _{x} & \frac{S}{2}(\Delta _{z} + 1) & 0 & \frac{S}{2}(\Delta _{z} + 1)(1+e^{-ik}) \\
\frac{S}{2}(\Delta _{z} + 1) & \frac{S}{2}(\Delta _{z} + 1) & h_{x} - 4S\Delta _{x} & \frac{S}{2}(\Delta _{z} + 1) & \frac{S}{2}(\Delta _{z} + 1) \\
\frac{S}{2}(\Delta _{z} + 1)(1+e^{ik}) & 0 & \frac{S}{2}(\Delta _{z} + 1) & h_{x} - 3S\Delta _{x} & 0 \\
0 & \frac{S}{2}(\Delta _{z} + 1)(1+e^{ik}) & \frac{S}{2}(\Delta _{z} + 1) & 0 & h_{x} - 3S\Delta _{x} \\
\end{bmatrix} .
\end{align}
\end{widetext}

The row vector in Eq.~(\ref{eq_vector3}) only contains boson creation operators and the Bogoliubov transformation are applied to obtain the eigenvalues $\varepsilon (k)$ of the dynamical matrix.


\begin{figure*}
\centering
\includegraphics[width=0.37\linewidth]{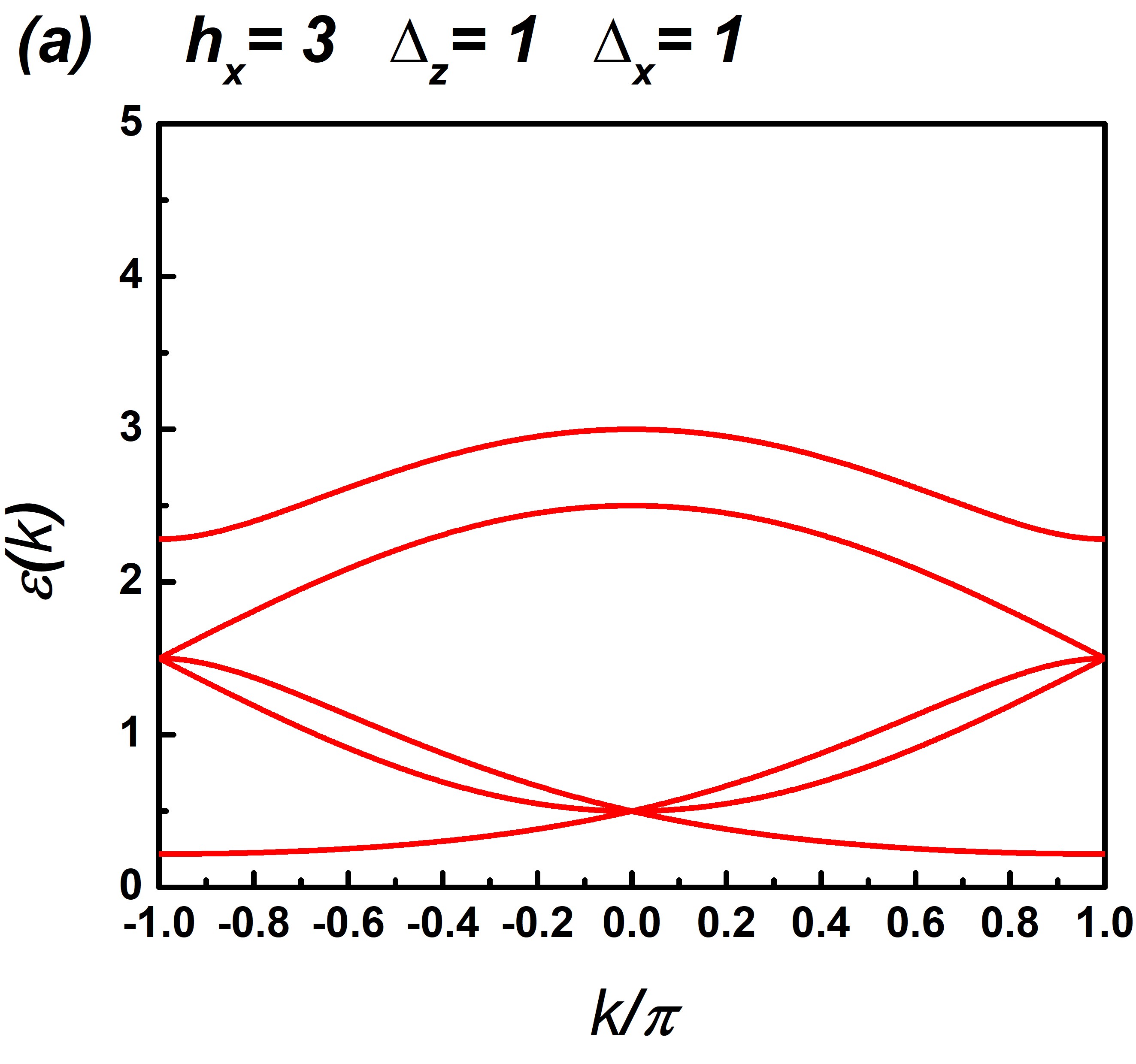}
\includegraphics[width=0.37\linewidth]{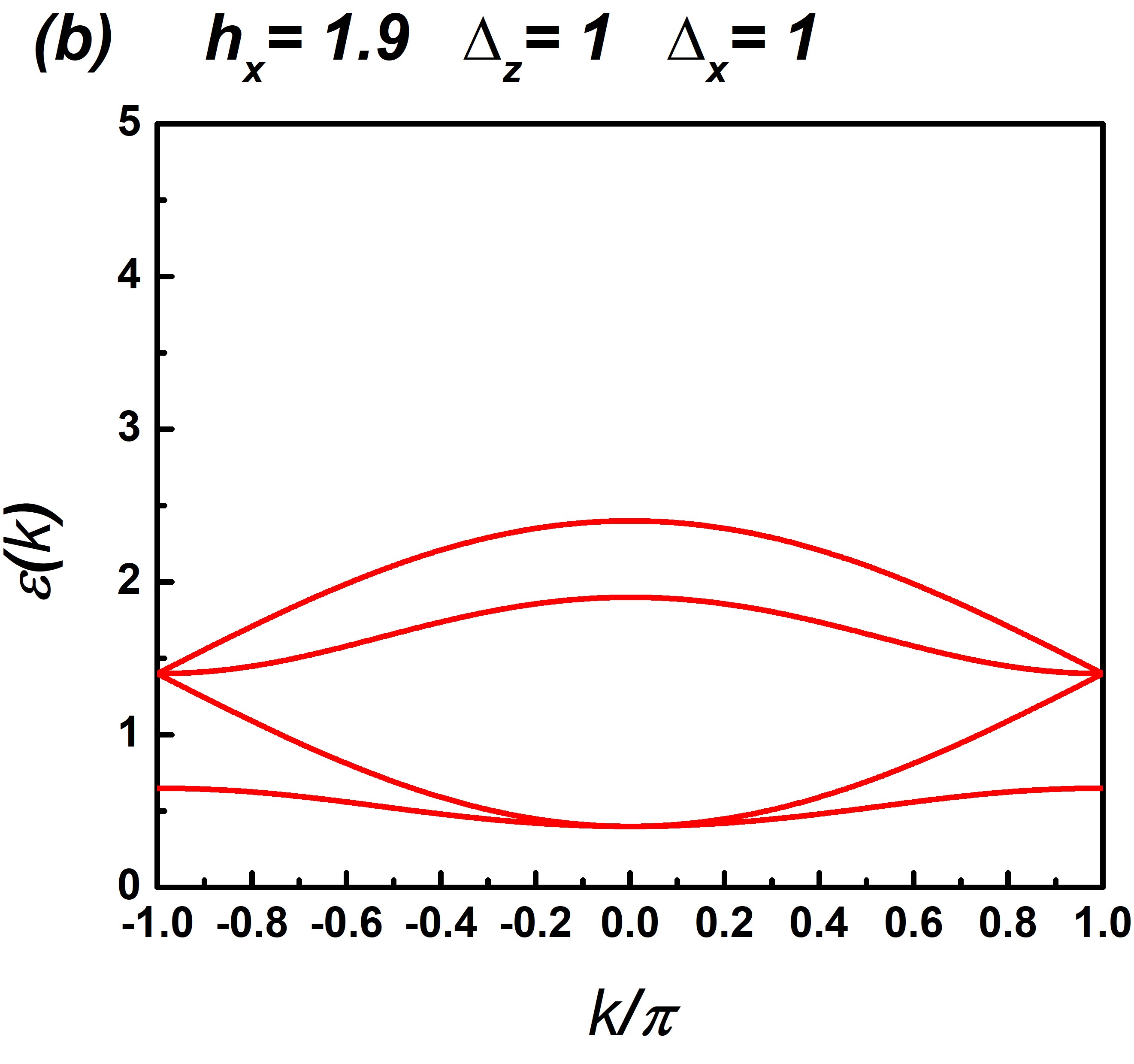}
\includegraphics[width=0.37\linewidth]{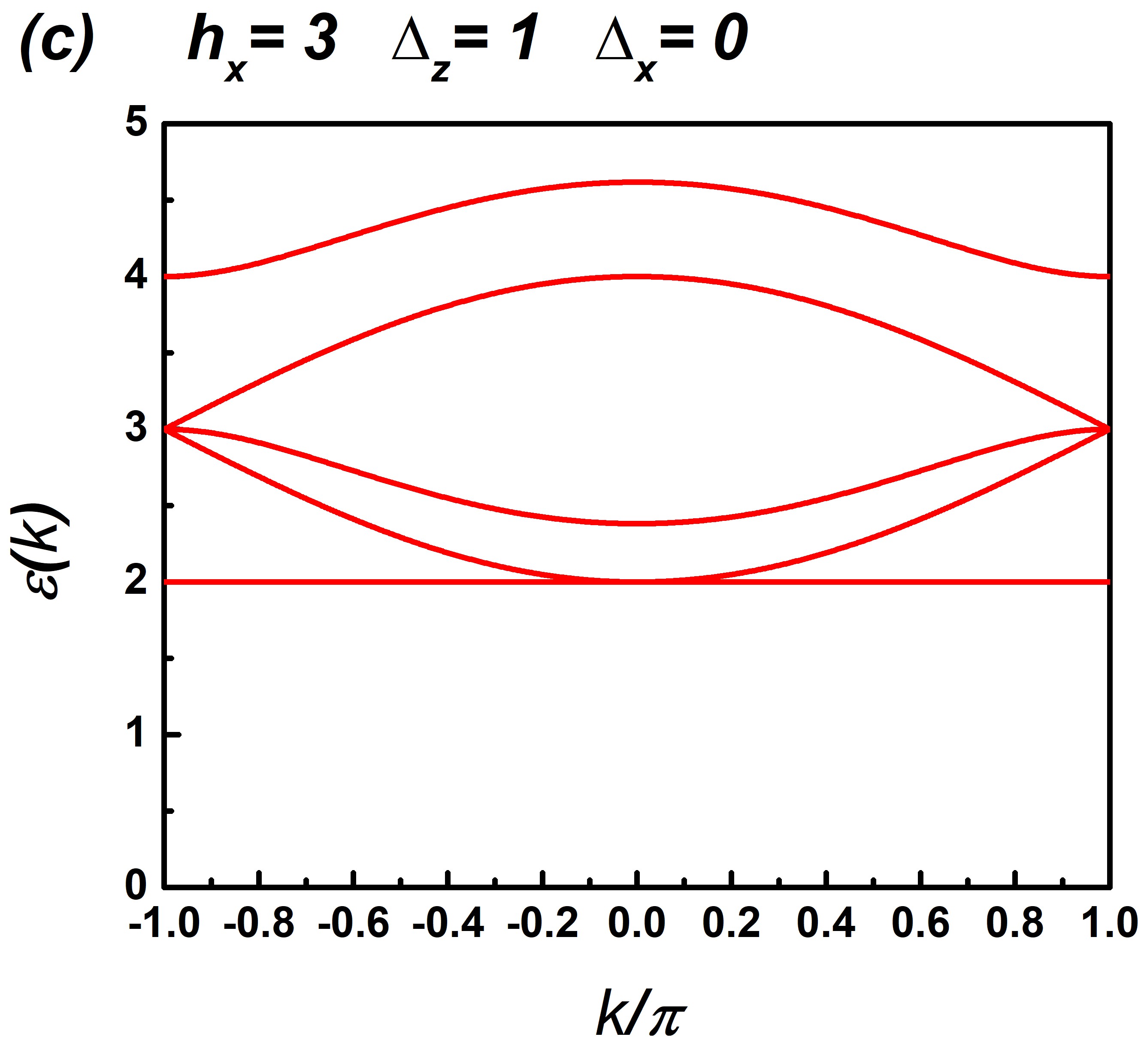}
\includegraphics[width=0.37\linewidth]{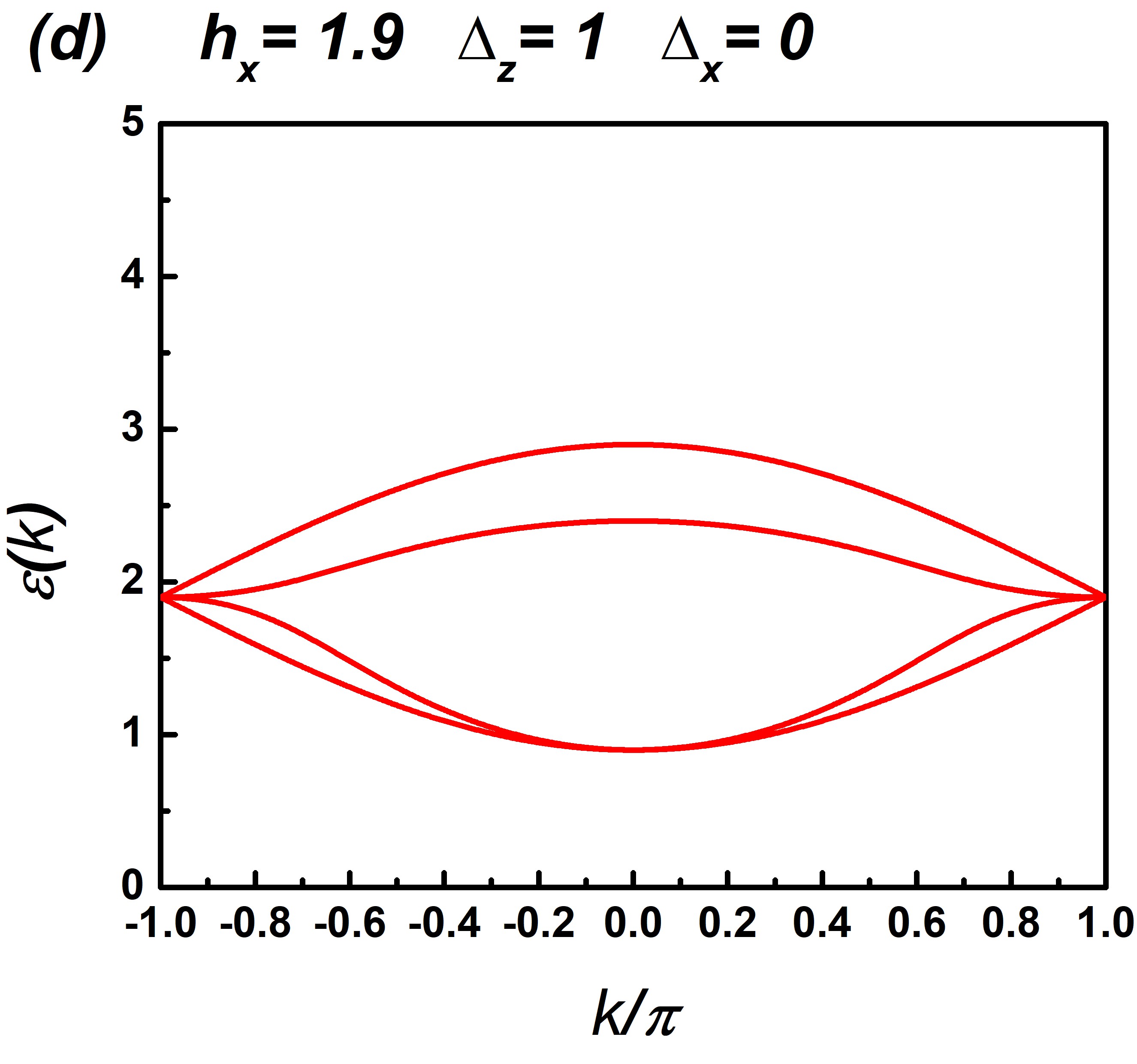}
\includegraphics[width=0.37\linewidth]{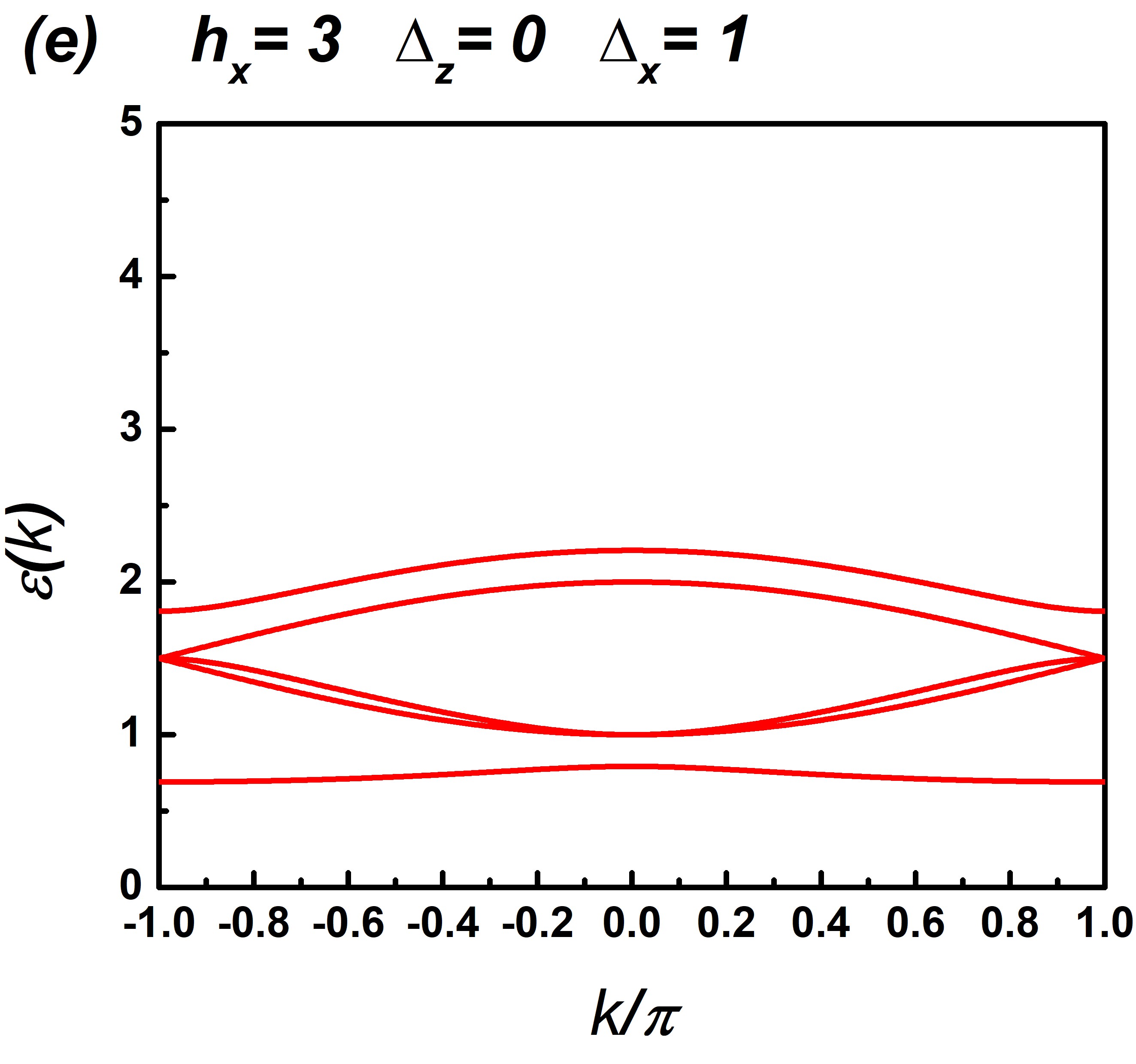}
\includegraphics[width=0.37\linewidth]{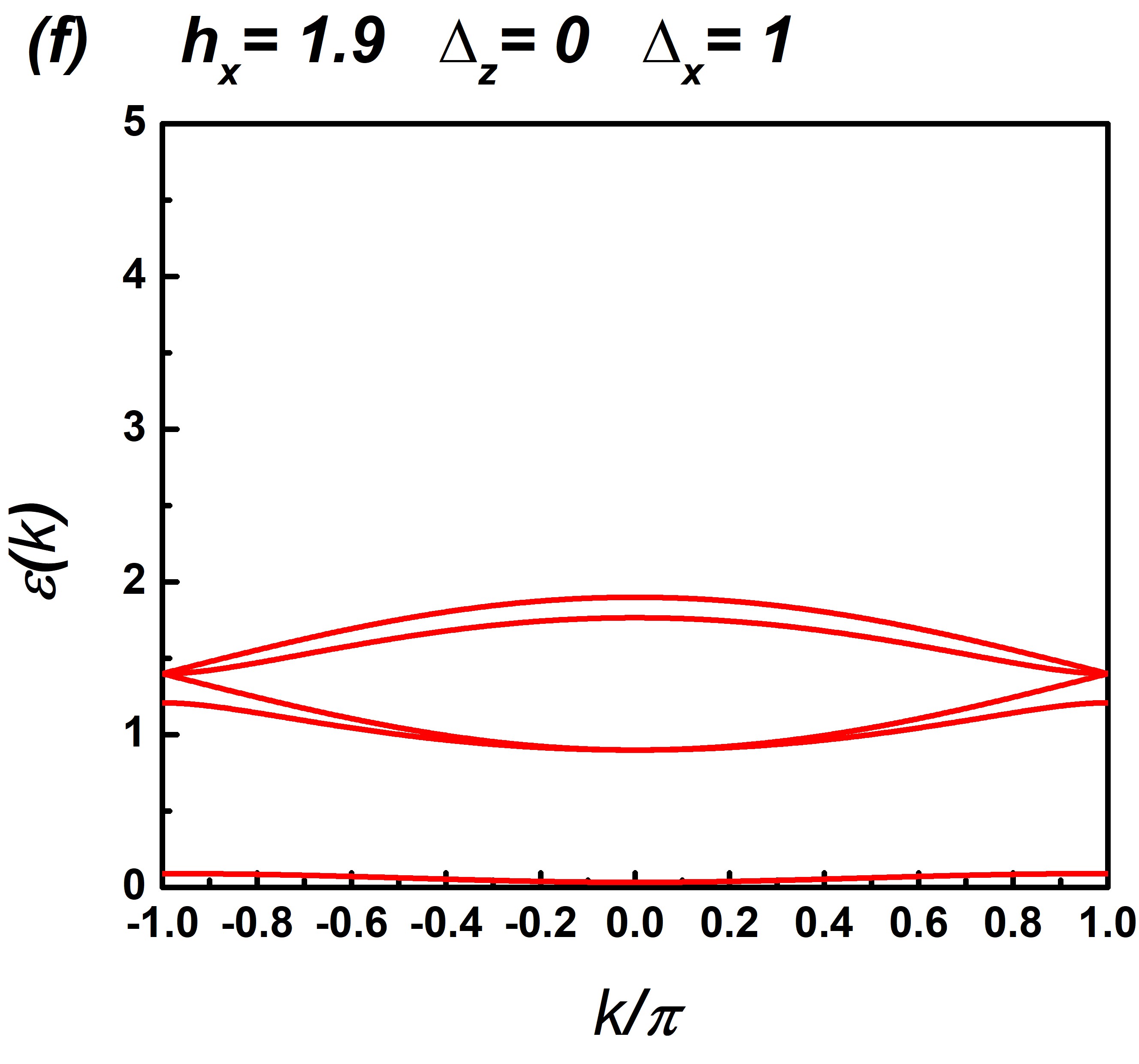}
\caption{The band structure for various $\Delta _{x}$ and $\Delta _{z}$. Panels (a), (c), and (e) are obtained for the polarized state using $h _{x} = 3$. Panels (b), (d), and (f) are obtained for the 0.6 plateau state using $h _{x} = 1.9$.}
\label{Fig_spinwave}
\end{figure*}

For the fully polarized state we choose $h_{x}=3$ as suggested by the DMRG results, and for the 0.6 plateau state we use an intermediate $h_{x}=1.9$ to obtain the magnon bands. As shown in Fig.~\ref{Fig_spinwave}(a), for the fully polarized state the magnon excitations are dispersive with a finite gap in the isotropic limit. The magnon gap is trivial because it can be directly tuned by the magnetic field $h_{x}$. When $\Delta _{x}$ is set to 0, the lowest band becomes flat while other bands remain dispersive, as shown in Fig.~\ref{Fig_spinwave}(c). The flat band can be interpreted as localized magnon excitations in real space~\cite{derzhko2007universal,huber2010bose}. Similar to the localized magnon state proposed in 2D kagome Heisenberg model~\cite{schmidt2006linear,schnack2020magnon}, the magnons with the same energy can contribute to a sudden drop from the saturation value~\cite{schulenburg2002macroscopic,zhitomirsky2004exact}. In the Appendix, we show more details of the numerical results of the magnetization curve for different $\Delta _{x}$, where the sudden drop below saturation and the emergent 0.8 plateau at $\Delta _{x}=0$ are consistent with Ref.~\cite{schulenburg2002macroscopic}. In contrast, when $\Delta _{z}$ is 0 while keeping $\Delta _{x}=1$, all bandwidths become smaller but finite, as shown in Fig.~\ref{Fig_spinwave}(e). 

Furthermore, we show the magnon excitations of the 0.6 plateau state for various $\Delta _{x}$ and $\Delta _{z}$. In the isotropic limit shown in Fig.~\ref{Fig_spinwave}(b), the bandwidth of the lowest band is finite. For $\Delta _{x}=0$ and $\Delta _{z}=1$ in Fig.~\ref{Fig_spinwave}(d), the band becomes significantly more dispersive than the isotropic limit. The dispersive band implies that the magnon excitations are non-local and can interact with each other. This can lead to the breaking of energy level degeneracy, which is consistent with the numerical results where the width of the 0.6 magnetization plateau becomes smaller; see more details of the magnetization curve near $M/M_{\text{sat}}=0.6$ for various $\Delta _{x}$ presented in the Appendix. For $\Delta _{x}=1$ and $\Delta _{z}=0$, the bandwidth remains small as shown in Fig.~\ref{Fig_spinwave}(f) and the 0.6 magnetization plateau is robust.


\begin{figure*}
\centering
\includegraphics[width=0.37\linewidth]{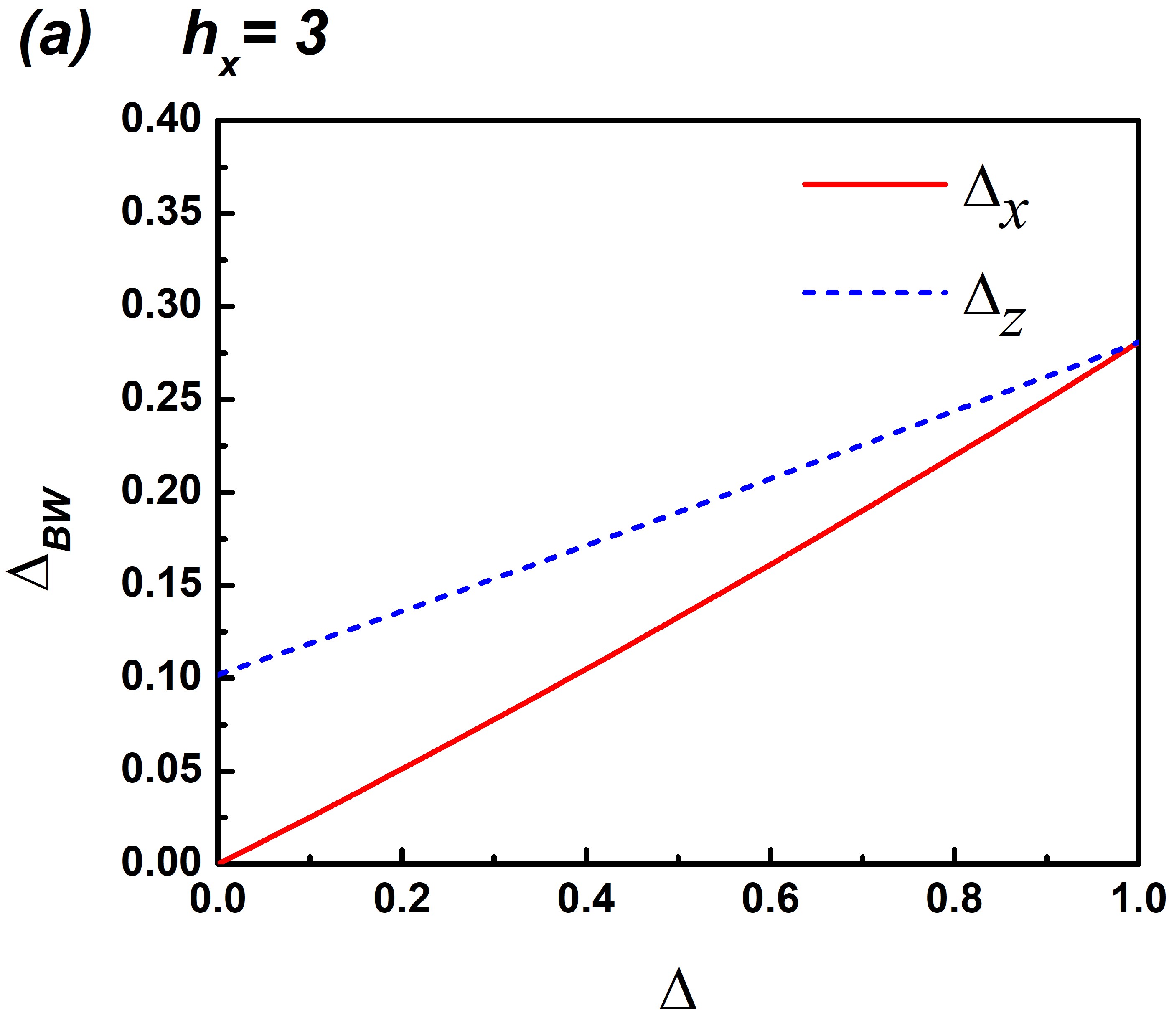}
\includegraphics[width=0.37\linewidth]{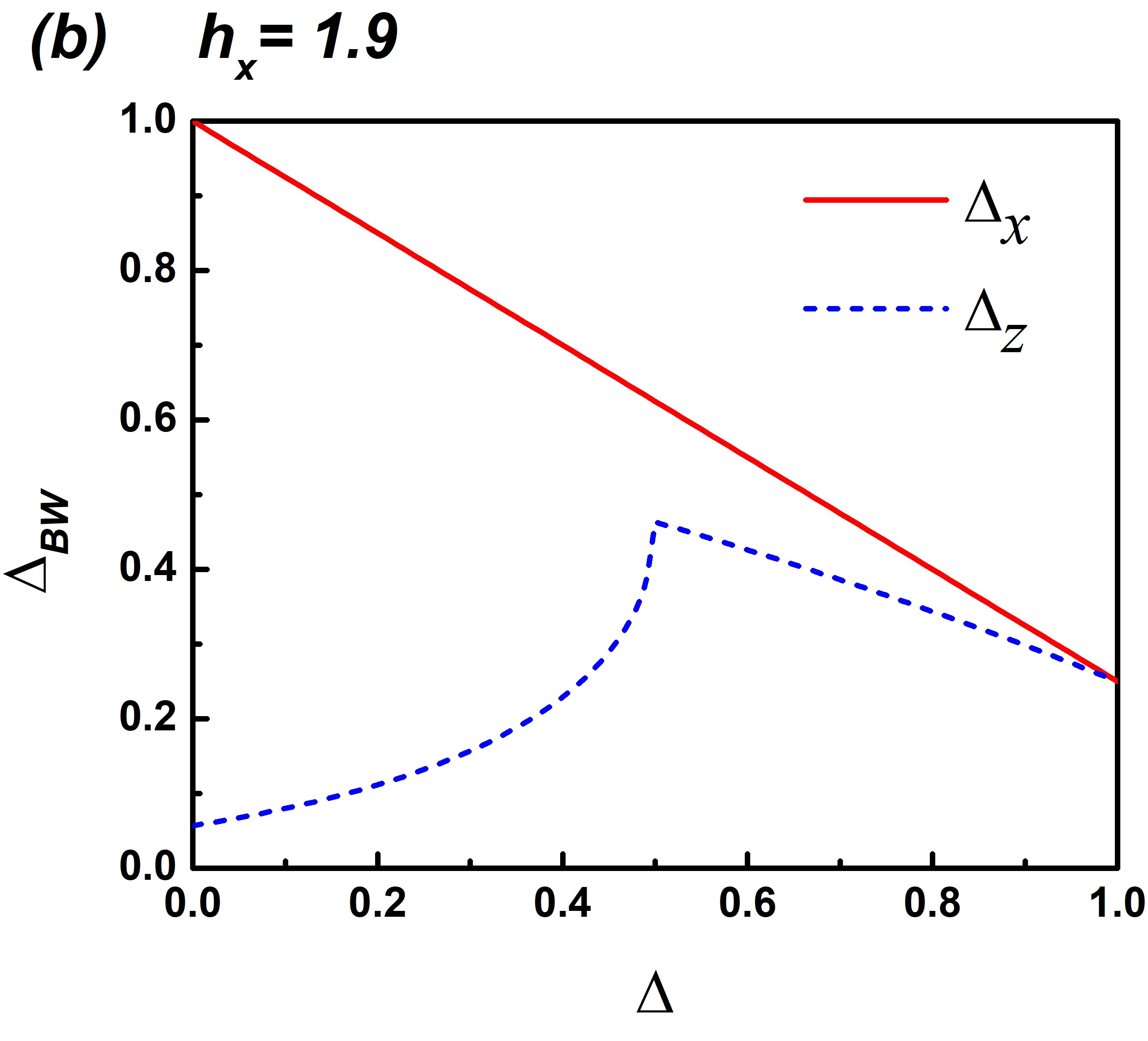}
\caption{The bandwidth for different $\Delta _{z}$ fixing $\Delta _{x} = 1$ and different $\Delta _{x}$ fixing $\Delta _{z} = 1$. Panel (a) is obtained for the polarized state using $h _{x} = 3$. Panel (b) is obtained for the 0.6 plateau state using $h _{x} = 1.9$.}
\label{Fig_BW}
\end{figure*}

To further characterize the flatness of the lowest band for various $\Delta _{x}$ and $\Delta _{z}$, we obtain the bandwidth as a function of the anisotropy. As shown in Fig.~\ref{Fig_BW}(a), the bandwidth corresponding to the polarized state decreases smoothly from 0.28 to 0 as $\Delta _{x}$ decreases from 1 to 0, indicating a completely flat band at $\Delta _{x}=0$. The bandwidth also decreases with decreasing $\Delta _{z}$ but remains finite for $\Delta _{z}=0$. For the magnon bands that correspond to the 0.6 plateau state, we find that the bandwidth increases monotonically with the decrease of $\Delta _{x}$, as shown in Fig.~\ref{Fig_BW}(b). As $\Delta _{z}$ decreases from the isotropic limit fixing $\Delta _{x}=1$, the bandwidth first has a slight increase, then decreases to 0.057 at $\Delta _{z}=0$. Our results show that with anisotropic interactions along the magnetic field, the band becomes more dispersive for the 0.6 plateau state while the bandwidth remains small but finite for a strong anisotropic interaction along directions perpendicular to the field, and the small bandwidth may be related to the stability of the plateau. For the polarized state the anisotropic interactions along different directions always result in a smaller bandwidth, and the band becomes completely flat at $\Delta _{x}=0$.

\section{Discussion and Summary}
\label{Summary}
We study the kagome strip chain Heisenberg model using both numerical DMRG methods and linear spin wave theory. We find that when the magnetic field is applied in the $x$ direction, the 0.6 magnetization plateau is stable against anisotropic interactions in the $z$ direction, but the plateau becomes much smaller with anisotropic interactions in the same $x$ direction. Spin wave analysis of the 0.6 plateau state shows that the bandwidth of the lowest magnon band remains small but finite for various $\Delta _{z}$, but the bandwidth increases as $\Delta _{x}$ decreases and the band becomes more dispersive. The dispersive band corresponds to the non-local magnon excitations and the magnons can interact with each other, which leads to the breaking of energy level degeneracy.

The results of the anisotropic effects could be tested experimentally through magnetization measurements under a magnetic field that is either parallel or perpendicular to the kagome strip chain~\cite{tang2016synthesis,zhang2020molybdate}. Although our study is restricted in one dimension, the geometry of the kagome strip chain and the 2D kagome lattice share some similarity. The results might have implications on the 2D kagome lattice materials that exhibit plateaus with magnetic field in a certain direction but no distinct plateau in the other direction~\cite{goto2017ising}, where the anisotropic g factors that result from spin-orbit couplings can contribute to the anisotropic Heisenberg interactions~\cite{rogado2003beta,amemiya2009partial,amemiya2012ferromagnetism,mokhtari20241}. Upon tuning the interaction strength of the lower legs below a critical value of $J_{d}=0.6$, the Hamiltonian of the kagome strip chain becomes dominated by one frustrated spin chain and one single-leg spin chain. These findings might be examined by replacing certain elements in the kagome strip chain materials to create different inhomogeneous couplings.
In addition to the material realization of kagome strip chain, our results may be tested using the transmon qubit technology by IBM~\cite{gambetta2017building, kjaergaard2020superconducting}, which can simulate Heisenberg models in a magnetic field.


\begin{acknowledgments}
This work was carried out under the auspices of the U.S. Department of Energy (DOE) National Nuclear Security Administration (NNSA) under Contract No. 89233218CNA000001.
It was supported by the online REU (O-REU) Program at Texas A\&M University  (C.B.). Y.H. was supported by the LANL LDRD Program and Center for Integrated Nanotechnologies, a DOE BES user facility, in partnership with the LANL Institutional Computing Program for computational resources. J.-X.Z. was supported by Quantum Science Center, a U.S. DOE Office of Science Quantum Information Science Research Center.
\end{acknowledgments}

%
%

\appendix
\setcounter{secnumdepth}{0}
\section{APPENDIX: ADDITIONAL NUMERICAL RESULTS NEAR PLATEAUS AND SATURATION}
\label{Apendix_saturation}

\begin{figure*}
\centering
\includegraphics[width=0.32\linewidth]{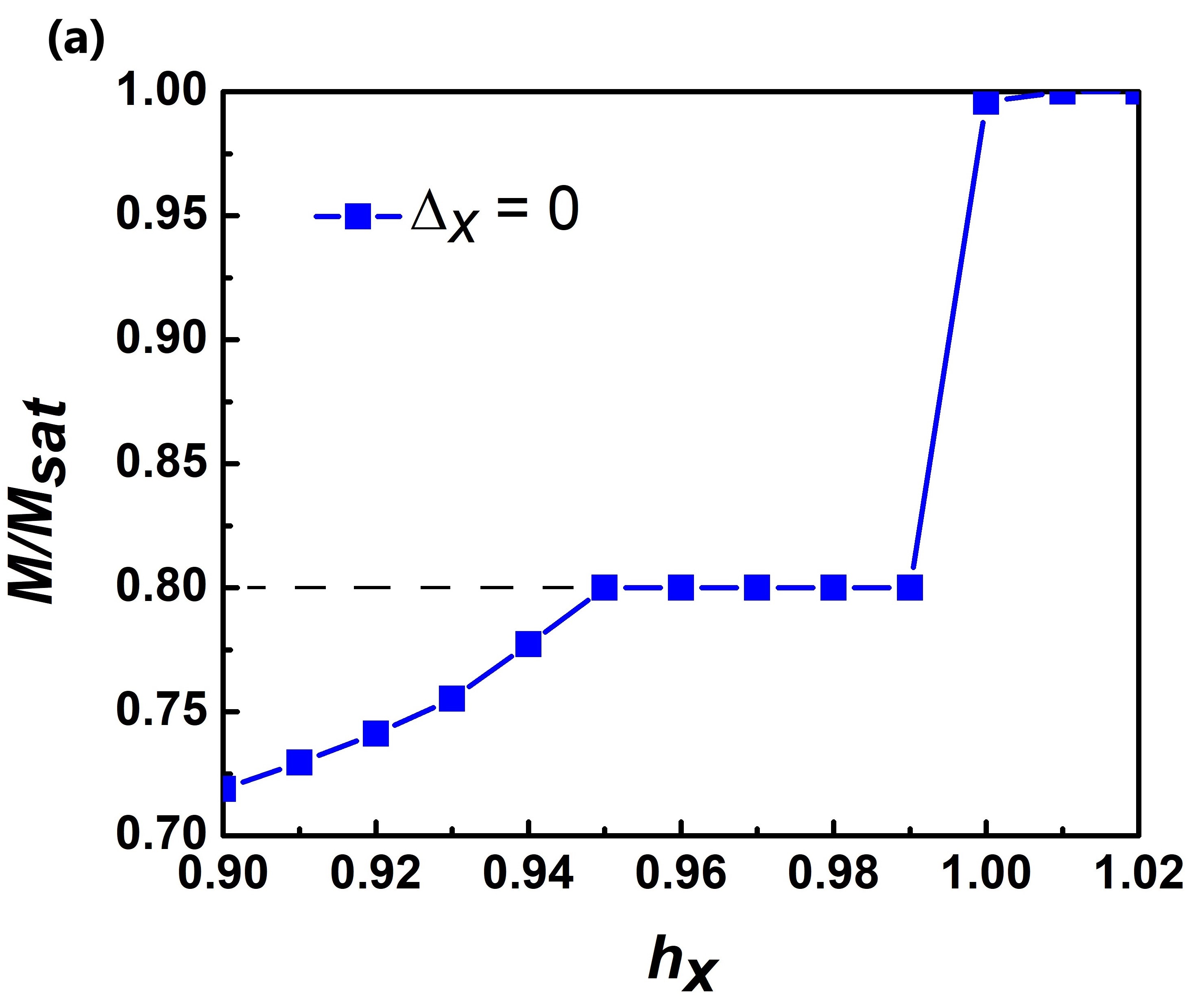}
\includegraphics[width=0.32\linewidth]{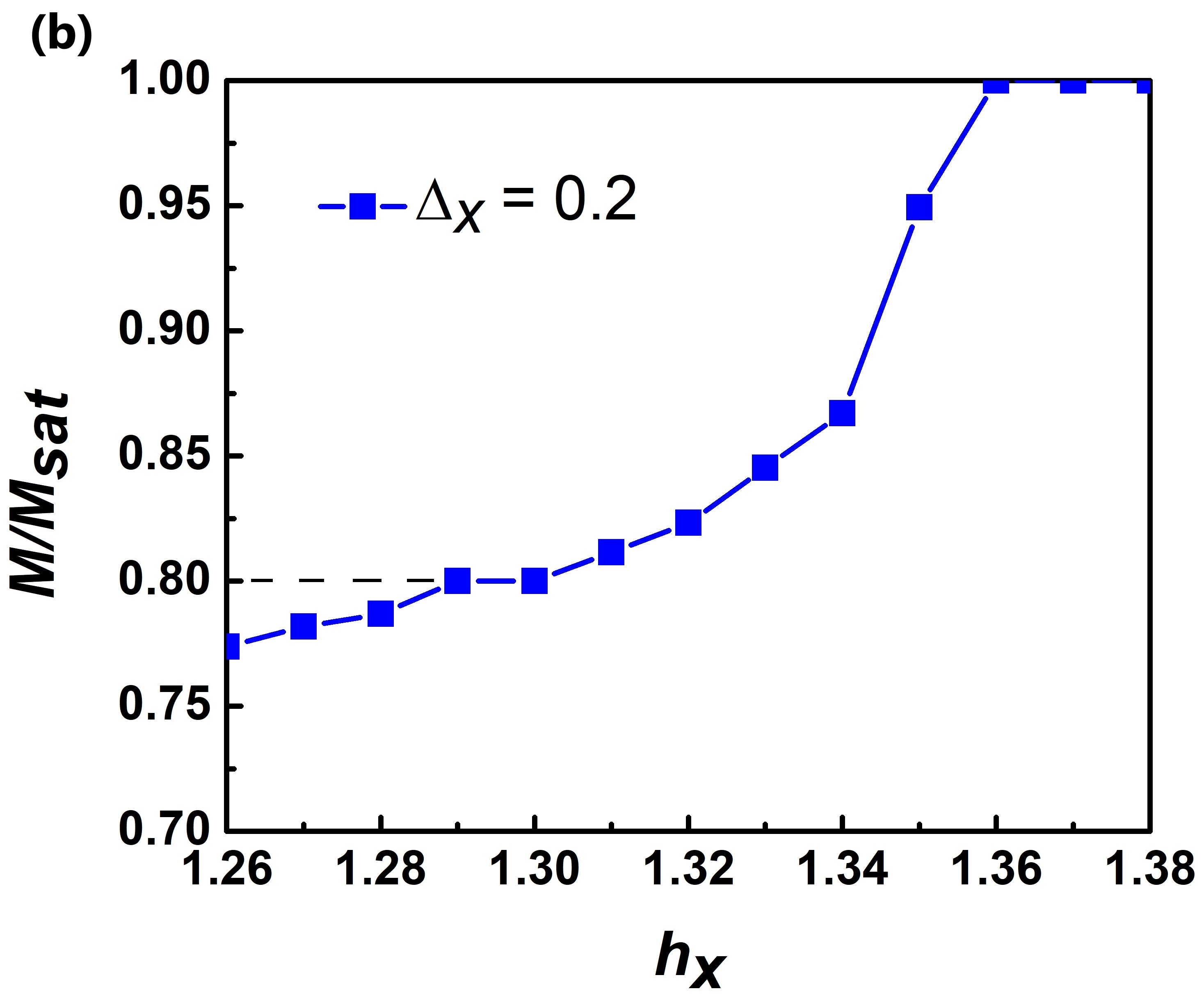}
\includegraphics[width=0.32\linewidth]{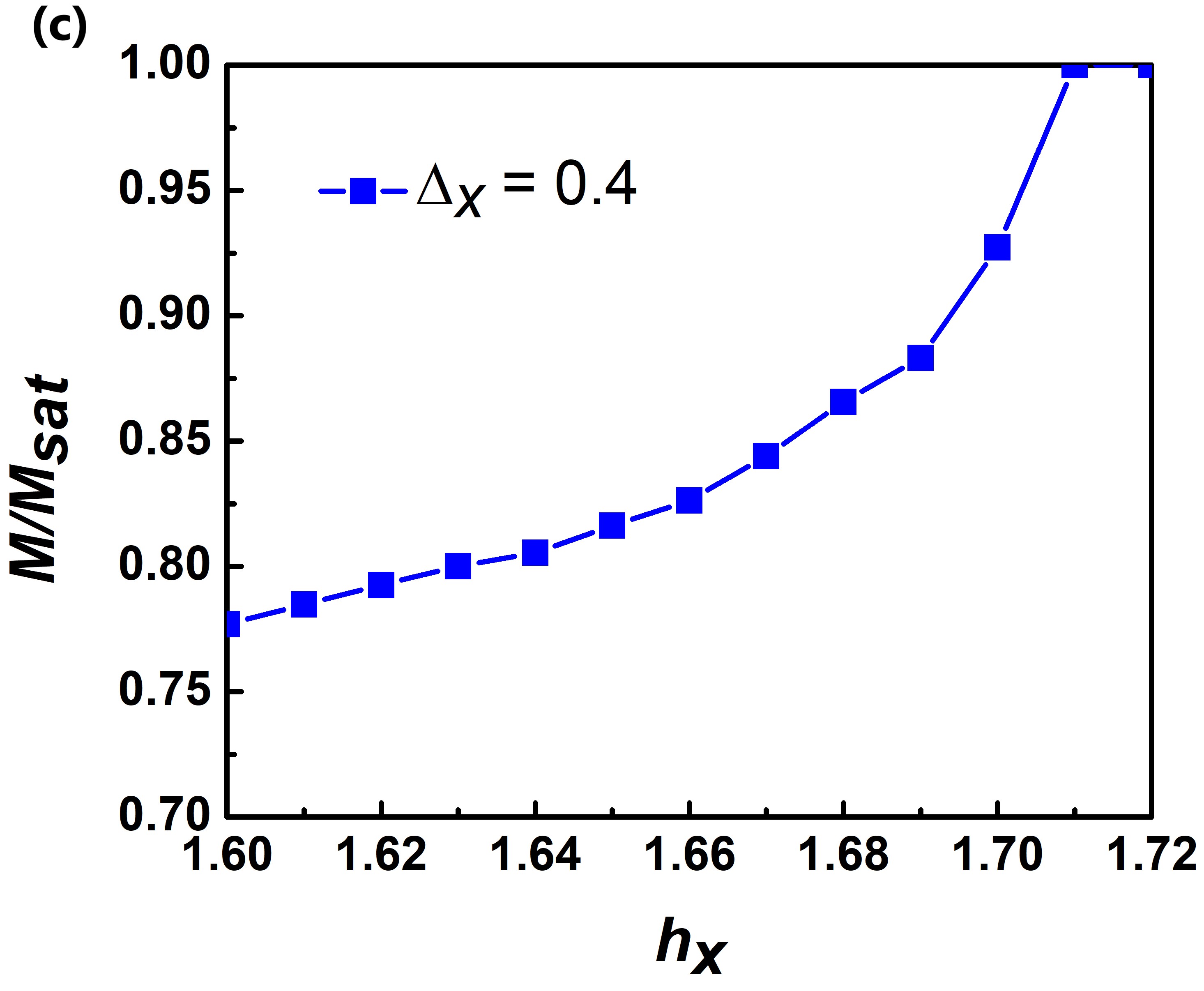}
\includegraphics[width=0.32\linewidth]{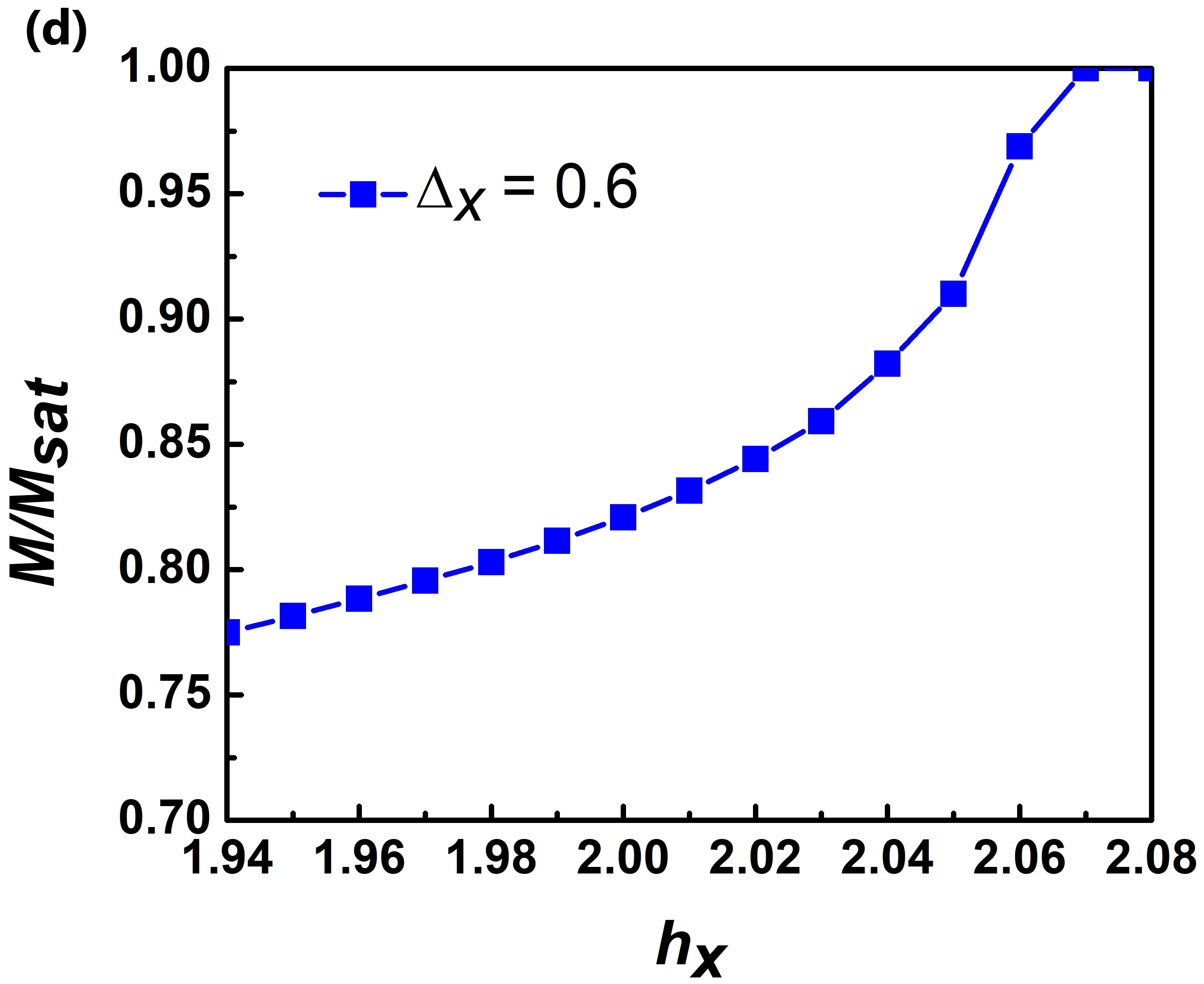}
\includegraphics[width=0.32\linewidth]{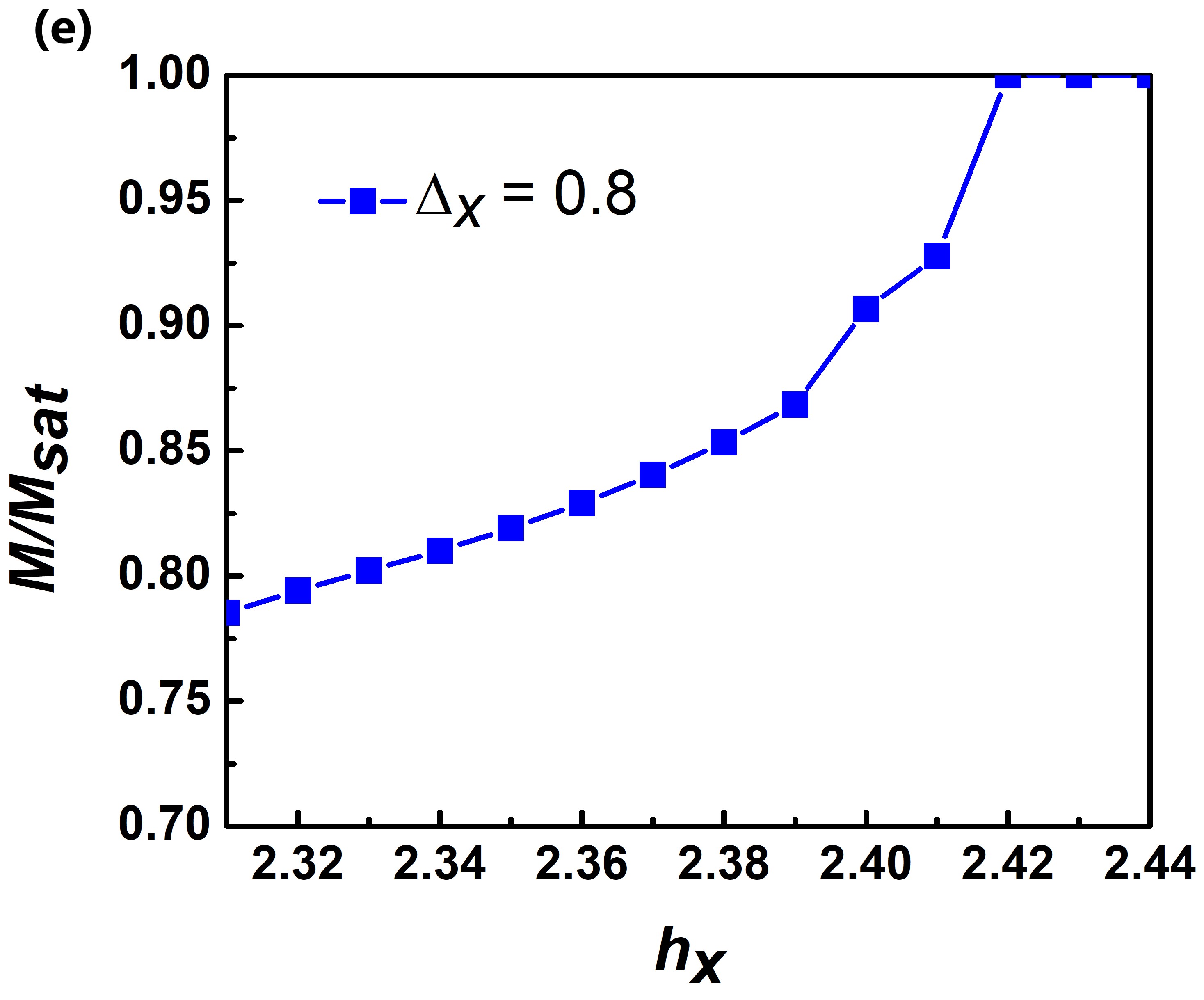}
\includegraphics[width=0.32\linewidth]{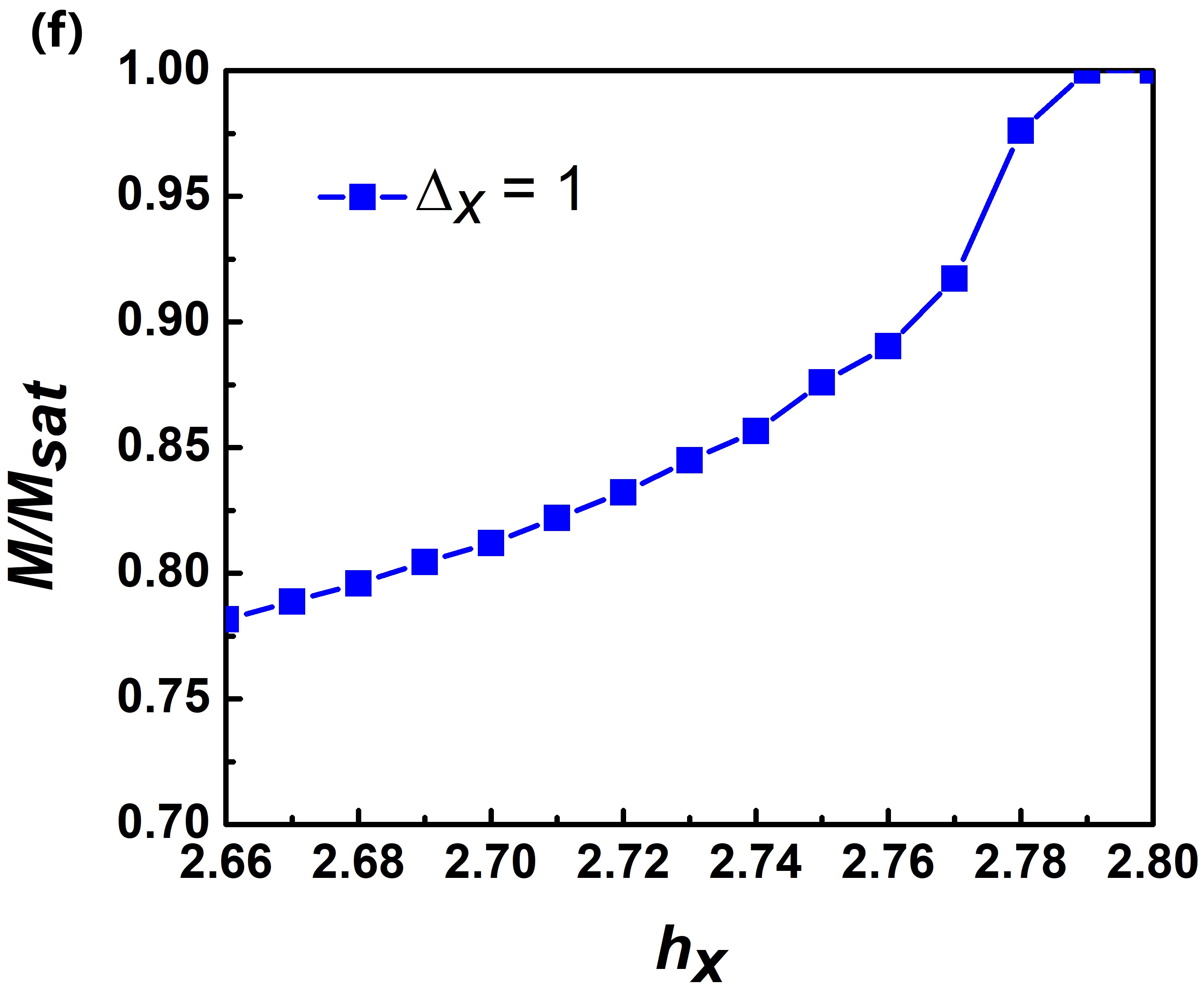}
\caption{\label{FigS:saturation} The DMRG results of the magnetization curve near saturation for various $\Delta _{x}$ at $\Delta _{z}=1$. The 0.8 plateau is indicated by the dashed line in panel (a) and (b). We set $J_{d}=J_{u} = J_{h} =1$ and use 6 - 12 unit cells with up to 400 bond dimensions to ensure the convergent results.}
\end{figure*}

\begin{figure*}
\centering
\includegraphics[width=0.32\linewidth]{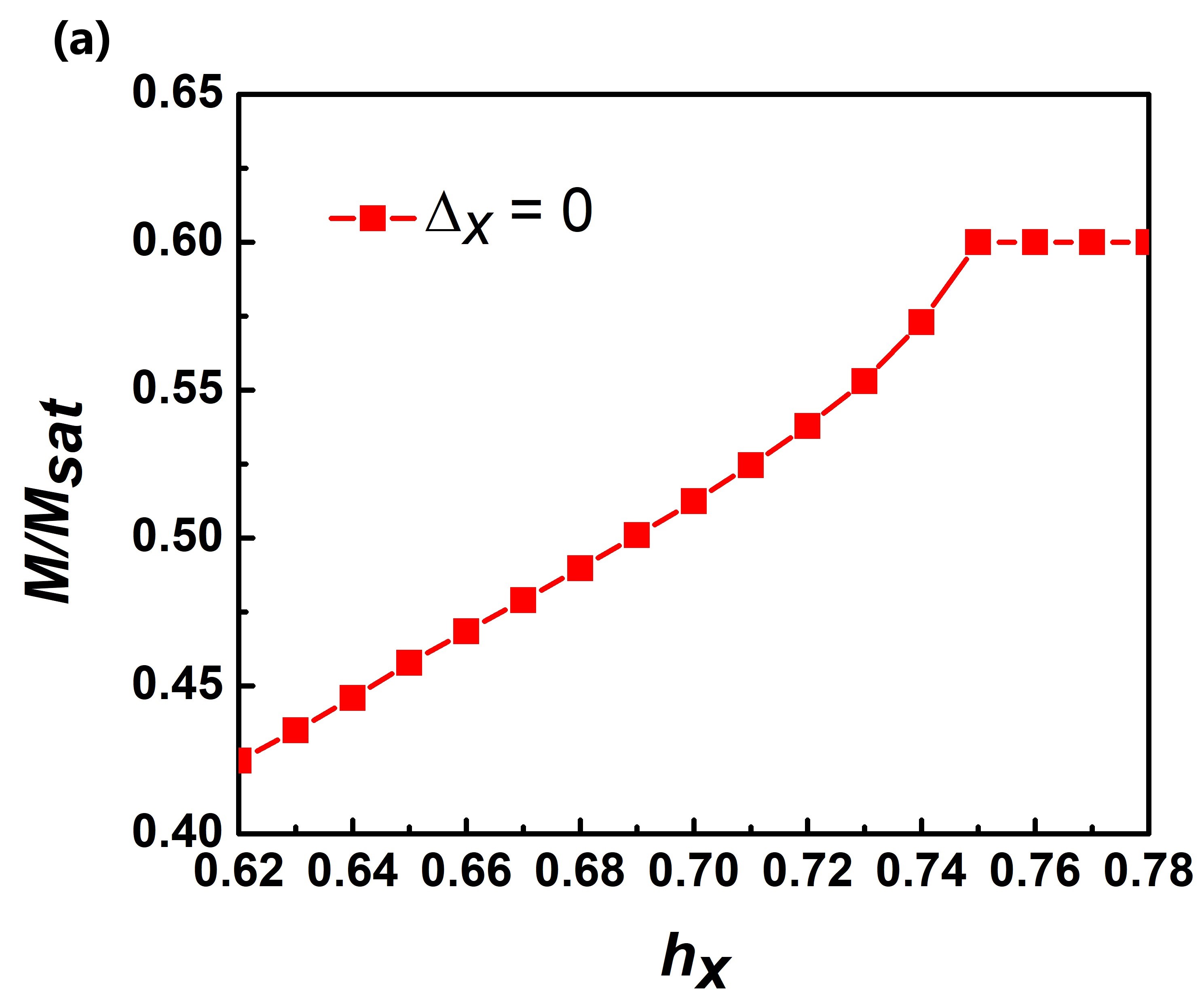}
\includegraphics[width=0.32\linewidth]{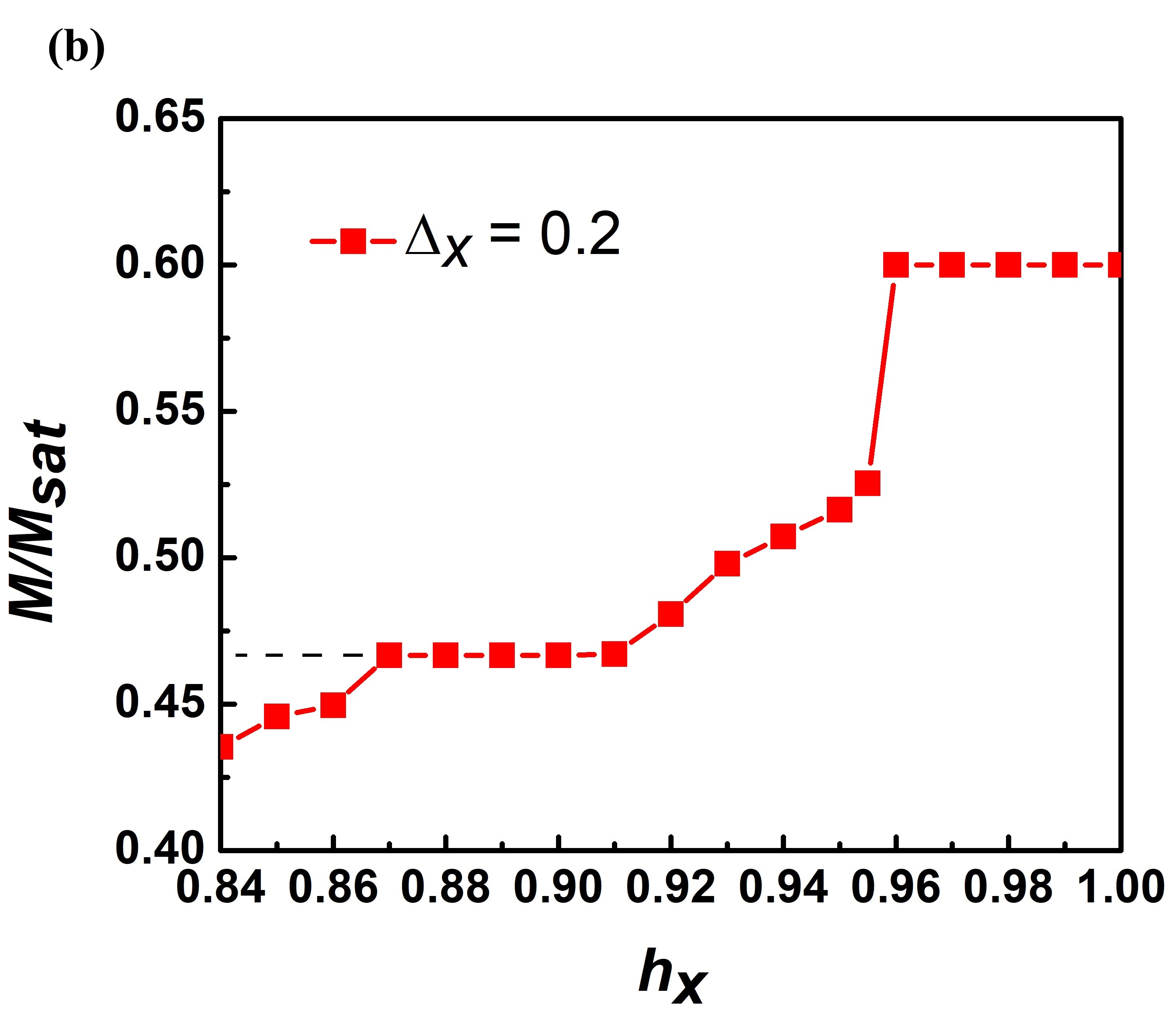}
\includegraphics[width=0.32\linewidth]{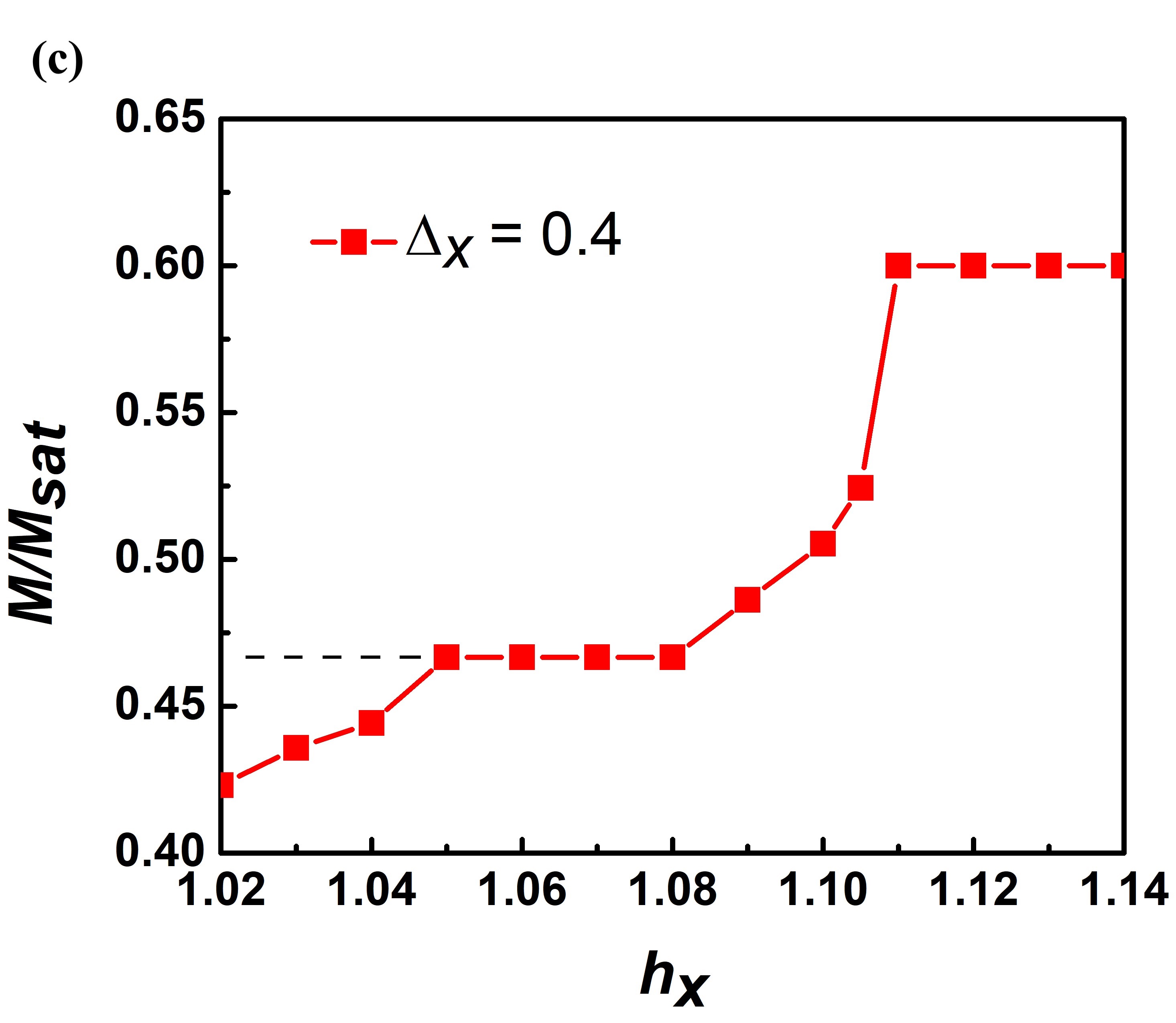}
\includegraphics[width=0.32\linewidth]{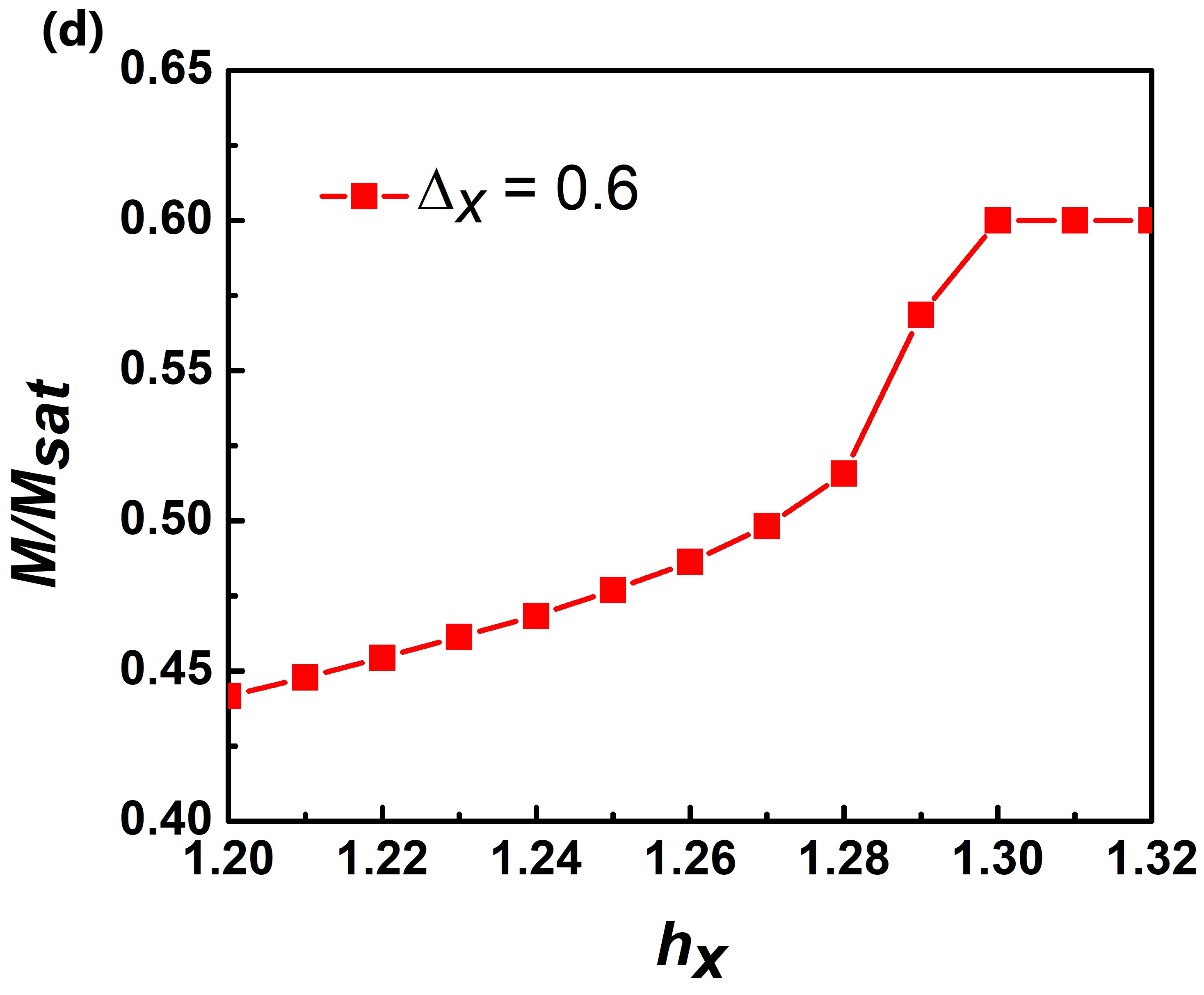}
\includegraphics[width=0.32\linewidth]{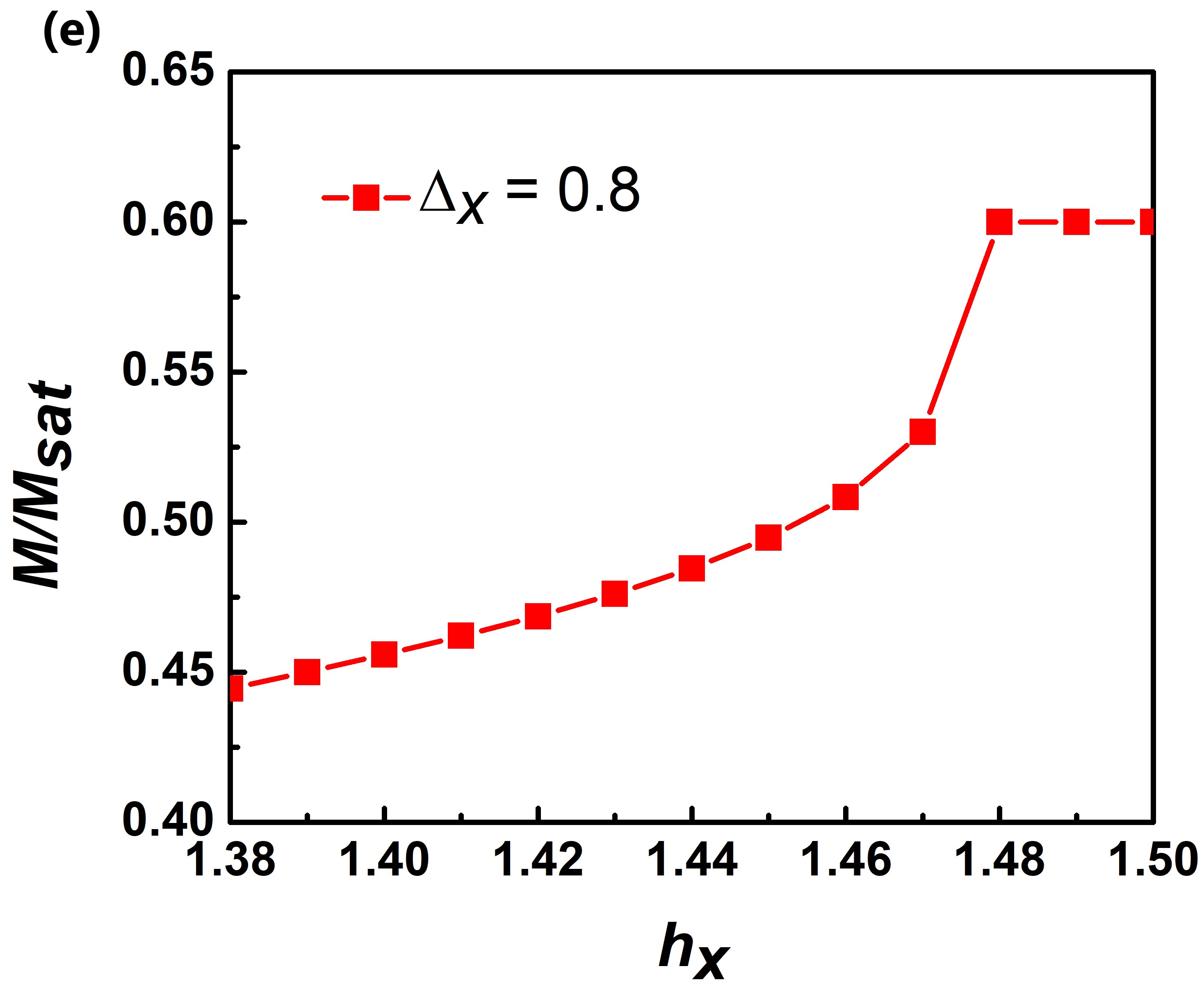}
\includegraphics[width=0.32\linewidth]{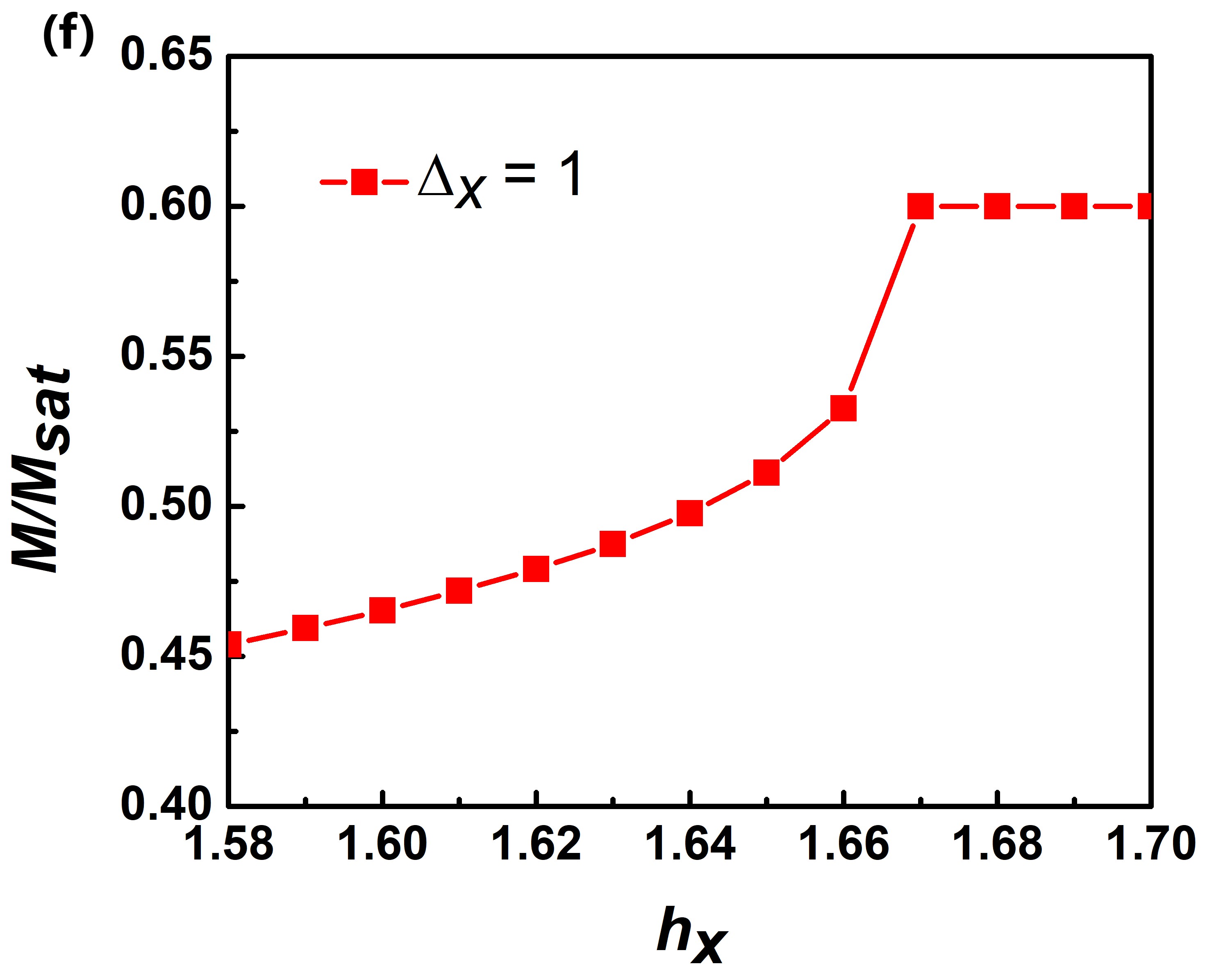}
\caption{\label{FigS:plateau} The DMRG results of the magnetization curve near $M/M_{\text{sat}}=0.6$ for various $\Delta _{x}$ at $\Delta _{z}=1$. The 7/15 plateau is indicated by the dashed line in panel (b) and (c). We set $J_{d}=J_{u} = J_{h} =1$ and use 6 - 12 unit cells with up to 400 bond dimensions to ensure the convergent results.}
\end{figure*}

\begin{figure}
\centering
\includegraphics[width=0.47\linewidth]{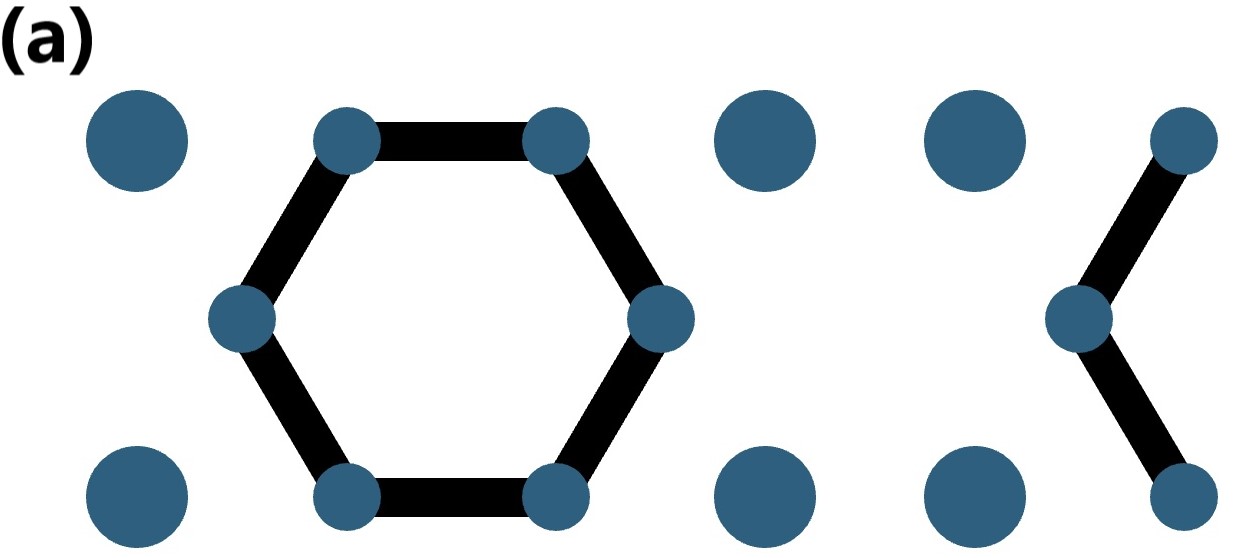}
\includegraphics[width=0.96\linewidth]{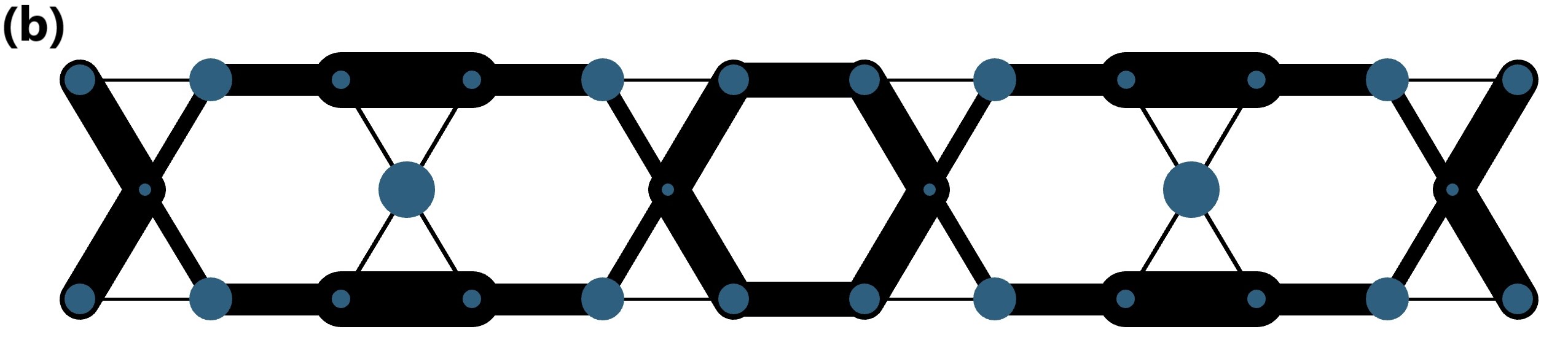}
\caption{\label{FigS:plateau715} The magnetic structure for the plateau at (a) $M/M_{\text{sat}}=0.8$ with $\Delta _{x}=0$ and (b) $M/M_{\text{sat}}=7/15$ with $\Delta _{x}=0.4$. The labeling is the same as the Fig.~\ref{Fig_struc} in the main text. For the 7/15 plateau we show $5 \times 6$ sites to characterize its periodic pattern. In panel (b), the $<S_{i}^{x}>$ at the second center site is 0.45, the $\chi(i,j)$ for the $J_{h}$ bond below the second center site is -0.46.}
\end{figure}

In order to examine the magnetization curve in details and explore the possible 0.8 plateau for various anisotropic interactions, we obtain the numerical DMRG results by varying $h_{x}$ with a smaller step of 0.01. As an example, we show the magnetization curve near saturation for various $\Delta _{x}$ at $\Delta _{z}=1$ in Figs.~\ref{FigS:saturation}(a) - (f). As shown in Figs.~\ref{FigS:saturation}(a) and (b), the 0.8 plateau exists for $\Delta _{x}<0.2$. The magnetic structure of the 0.8 plateau is given in Fig.~\ref{FigS:plateau715}(a) with a period of $2 \times 5$ sites. One localized magnon is identified in the hexagon and the rest of the spins remain polarized, which agrees with previous study~\cite{schulenburg2002macroscopic}. There is a sudden drop from saturation value for $\Delta _{x}=0$, which is consistent with a flat magnon band obtained from the linear spin wave analysis in the main text and also agrees with previous results using the exact diagonalization methods~\cite{schulenburg2002macroscopic}. As the bandwidth increases by tuning $\Delta _{x}$ to be larger, the magnetization curve becomes smooth and the 0.8 plateau vanishes. The macroscopic drop is not observed in the isotropic limit of $\Delta _{x}=1$ where the magnon bandwidth is finite, as shown in Fig.~\ref{FigS:saturation}(f). Starting from the 0.6 plateau, there is no sudden macroscopic drop in magnetization for all $\Delta _{x}$ as shown in Figs.~\ref{FigS:plateau}(a) - (f), which is consistent with a finite bandwidth. Interestingly, we identify the emergence of a new plateau at $M/M_{\text{sat}}=7/15$ for $0.2<\Delta _{x}<0.4$, as shown in Figs.~\ref{FigS:plateau}(b) and (c). The magnetic structure is identified with the period of $3 \times 5$ sites, as shown in Fig.~\ref{FigS:plateau715}(b). The 7/15 plateau also satisfies the Oshikawa-Yamanaka-Affleck condition since $n_{uc}p(M_{sat}-M)=4$. However, the emergence of this plateau cannot be explained by the spin wave analysis of the 0.6 plateau state, which may require further study. We notice that similar 7/15 plateau can be identified by tuning the coupling strengths at different bonds in the isotropic antiferromagnetic Heisenberg model on the kagome strip chain~\cite{morita2018magnetization,morita2021magnetic}.

\bibliography{Plateau_KS}

\end{document}